\documentclass[useAMS,usenatbib]{mn2e}

\usepackage{graphicx}

\usepackage{latexsym}

\usepackage{amsmath}

\usepackage{subfigure}

\usepackage{wasysym}

\title[The Theory of Photoevaporation]{On the theory of disc photoevaporation}

\author[J. E. Owen, C. J. Clarke \& B. Ercolano]{James E. Owen$^{1,2}$\thanks{E-mail: jowen@cita.utoronto.ca}, Cathie J. Clarke$^{1}$ \&
    Barbara Ercolano$^{3,4}$  \\ $^1$Institute of
    Astronomy, Madingley Road, Cambridge, CB3 0HA, UK\\ $^2$Canadian Institute for Theoretical Astrophysics, 60 St. George Street, Toronto, ON M5S 3H8, Canada\\ $^3$Excellence Cluster Universe, Boltzmannstr. 2, 85748 Garching, Germany\\ $^3$University Observatory,  Ludwig-Maximillans University Munich, Scheinerstr. 1, 81679 Munich, Germany}

\begin{document}

\pagerange{\pageref{firstpage}--\pageref{lastpage}} \pubyear{2002}

\newcommand{\dd}{\textrm{d}}
\newcommand{\Tesc}{T_{\textrm{esc}}}

\newcommand{\rin}{R_\textrm{\tiny{in}}}

\newcommand{\OO}{\mathcal{O}}
\newcommand{\ngg}{\gg \hspace{-0.36cm} /\hspace{0.2cm}}
\newcommand{\msun}{M$_\odot$}

\newcommand{\msunyr}{M$_\odot$~yr$^{-1}$}

\maketitle

\def\lsun{{\rm L_{\odot}}}

\def\rsun{{\rm R_{\odot}}}
\def\rl{{R_{\rm L}}}

\label{firstpage}

\def\mnras{MNRAS}

\def\apj{ApJ}

\def\aap{A\&A}

\def\apjl{ApJL}

\def\apjs{ApJS}

\def\araa{ARA\&A}

\begin{abstract}
We discuss a hydrodynamical model for the dispersal of protoplanetary discs around young, low mass ($<1.5$ \msun) stars by photoevaporation from the central object's energetic radiation, which considers the far-ultraviolet as well as the X-ray component of the radiation field. We present analytical scaling relations and derive estimates for the total mass-loss rates, as well as discussing the existence of similarity solutions for flows from primordial discs and discs with inner holes. Furthermore, we perform numerical calculations, which span a wide range of parameter space and allow us to provide accurate scalings of the mass-loss rates with the physical parameters of the systems (X-ray and FUV luminosity, stellar mass, disc mass, disc temperature and inner hole radius).

The model suggest that the X-ray component dominates the photoevaporative mass-loss rates from the inner disc. The mass-loss rates have values in the range from 10$^{-11}$ to 10$^{-7}$ \msunyr and scale linearly with X-ray luminosity, with only a weak dependence on the other parameters explored. However, in the case of high FUV to X-ray ($L_{\rm FUV}/L_X>100$) luminosity ratios, the FUV constricts the X-ray flow and may dominate the mass-loss.

Simulations of low mass discs with inner holes demonstrate a further stage of disc clearing, which we call `thermal sweeping'.  This process occurs when the mid-plane pressure drops to sufficiently low values. At this stage a bound, warm, X-ray heated region becomes sufficiently large and unstable, such that the remaining disc material is cleared on approximately dynamical time-scales. This process significantly reduces the time taken to clear the outer regions of the disc, resulting in an expected transition disc population that will be dominated by accreting objects, as indicated by recent observations.

\end{abstract}


\section{Introduction}

The evolution and dispersal of protoplanetary discs is an important
step in the star and planet formation process. Protoplanetary discs provide the material from which planets
form, implying that the initial evolution of newly formed planets is coupled to the
properties of the parent disc. Moreover the disc dispersal time-scale
ultimately sets the time in which (gas) planets must form. 

It is observationally well established that at an age of 1~Myr most
    stars are surrounded by discs that are optically thick at
    infra-red (IR) wavelengths, which have in most cases disappeared by
    typical ages of 10~Myr (e.g. Haisch et
    al. 2001, Mamajek 2009). Optically thick discs around young stars evolve in a manner that
    is consistent with  `standard' viscous evolution theory, where the disc
    diagnostics decline with time in an approximately power law 
    fashion (Hartmann et
    al. 1998). However, the final evolution from disc-bearing (primordial)
    to disc-less status appears to be much more rapid than viscous
    theory predicts (e.g Luhman et al. 2010;
    Muzerolle et al. 2010; Ercolano et al. 2011) and to occur 
    over the entire range of disc radii probed by IR
    observations. Furthermore, given that the sub-millimetre observations show that
    most non-accreting stars (WTTs) are devoid of emission out to
    several hundreds of AU (Duvert et al. 2000, Andrews \& Williams 2005, 2007),
    the clearing must be correlated across all disc radii from 1AU to 100AU.

Further clues regarding the mechanism for clearing discs are 
provided by objects which have been identified as `transition' discs. These 
are discs which show a deficit in opacity at near-IR (NIR) wavelengths compared
to a standard optically thick `primordial'  disc, but are consistent
with standard optically thick discs at Mid-IR/Far-IR (MIR/FIR)
    wavelengths; furthermore some of these objects show small NIR
    excesses above the photosphere (not consistent with a primordial
    disc) have been labelled as `pre-transition' discs (e.g. Espaillat
    et al. 2010; Furlan et al. 2011). Such
spectral characteristics are  
most readily explained in terms of a hole or gap in the dust in the inner
regions of the disc. However, there is no clear consensus as to the
origin of such structures   
and the debate is complicated by the
fact that they display a  wide variety of
other observational characteristics: for example some show evidence for
accretion
(e.g. Calvet et al. 2005, Najita et al. 2007,Hughes et al. 2009) while others 
do not (Cieza et al. 2010; Merin et al. 2010); there is also evidence in some
systems that the deficit in opacity at NIR wavelengths
does not necessarily preclude significant quantities of gas and small
    amounts of dust in the
inner disc (Espaillat et al. 2010) although this is not
always the case 
(Calvet et al. 2002). The frequency of these { `transition/pre-transition'} discs
compared to those of primordial discs suggest that approximately
10-20\% of discs are in this stage (Strom et al. 1989, Skrutskie et
al. 1990, Luhman et al. 2010, Ercolano et al. 2011a; Furlan et al. 2011). Assuming that
these objects are in transition from a primordial to  
a fully cleared state, this observation provides  two important
constraints: the transition time from primordial to cleared is
approximately 10\% of the discs total lifetime, and that the clearing
process proceeds from the inside out.   

While many theories have been developed in order to explain the
    observations of `transition' discs: grain growth (Dullemond \&
    Dominiki, 2005); photophoresis (Krasuss et al. 2007); MRI-driven
    winds (Suzuki \& Inutsuka, 2009), { only two mechanism have
    been proposed that can explain
    the observations of `transition' discs in the context of total
    (gas \& dust) 
    disc clearing (assuming that indeed all discs pass through a
    `transition' disc phase). These are} photoevaporation (e.g. Clarke et al. 2001) and
    planet formation (e.g. Armitage \& Hansen, 1999). Photoevaporation
    and planet formation are processes which compete to remove gas from
    the disc, and it is still a matter for debate whether one process
    dominates over the other. Ercolano \& Clarke (2010) discuss the
    implications of metallicity on both processes and Yasui et
    al. (2009) provide very preliminary observations of smaller disc
    frequencies in low metallicity regions in the extreme outer
    Galaxy, consistent with the predictions from X-ray
    photoevaporation models and arguing against planet formation as a
    main dispersal mechanism. 

Like any theory, the photoevaporation model has progressed through
    several stages of evolution, { with the original theories only
    including photoevaporation without viscous disc evolution (Shu et
    al. 1993; Hollenbach et al. 1994; Yorke \& Welz 1996; Richling \&
    Yorke 1997), or photoevaporation from
    external sources (Johnstone et al. 1998; Richling \& Yorke 1998;
    Richling \& Yorke 2000; Adams et al. 2004). It was} Clarke et
    al. (2001) who realised that
    mass-loss from the surface of a disc due to heating
    (photoevaporation), coupled with viscous evolution results in a
    gap opening in the inner disc when the photoevaporation rate and
    accretion rate become comparable. The inner disc then drains onto
    the central star on its own much faster local viscous
    time-scale. While the basic ideas have remained identical
    (photoevaporation and viscous process compete until
    photoevaporation eventually opens a gap in the inner disc which
    then rapidly drains on to the central star), the actual
    calculations of the photoevaporation rates have progressed
    greatly. The original model used semi-analytic mass-loss rates due
    to EUV radiation calculated by Hollenbach et al. (1994), while 
later the photoevaporative flow was modelled hydrodynamically  by Font
et al. (2004).  In these models the disc is mainly irradiated by a
diffuse field 
of recombination photons from the atmosphere of the inner
disc. Subsequently Alexander et al. (2006a) realised that, following
the 
draining of the inner disc,   the inner-edge of the hole is instead directly
exposed to the
radiation field from the star; this results in higher mass-loss 
rates which rapidly clear the remaining disc out to large radii (Alexander et al. 2006b). 

 Recently, a lot of attention has been paid to other high energy
 radiation fields; X-rays (Alexander et al. 2004, Ercolano et
 al. 2008,2009, Owen et al. 2010,2011b) and FUV (Gorti \& Hollenbach
 2008,2009; Gorti et al. 2009). These calculations have yielded much higher photoevaporation rates than
 the original EUV rate. {These higher mass-loss rates primarily  arise from significant mass-loss at larger radii
    in the disc ($R>$10 AU), due to  the ability of the FUV and
    X-rays to heat much larger columns of gas to of order the escape temperature}. While the central star certainly
 produce significant fluxes of FUV, EUV and X-rays (e.g. Alexander et al. 2005, Ingleby et
 al. 2011) it is important to determine which, or which combination of
 high energy rates are responsible for setting the photoevaporation
 rate.  

{ Some of the main observational tests of photoevaporation are
    based on observational statistics of disc populations as a
    function of time}. Owen et al. (2011b) showed that the X-ray
    photoevaporation model is consistent with the current diagnostics
    of disc evolution, and made some predictions about how
    these diagnostics should correlate with X-ray
    luminosity. Furthermore, Drake et al. (2009) reported that accretion
    rate measurements and X-ray observations in  Orion, point to a
    process of X-ray driven `photoevaporation-starved accretion',
    where higher mass-loss rates prevent material accreting onto the
    central star, leading to lower accretion rates. This process was
    theoretically demonstrated by Owen et al. (2011b)
    and is also likely to  explain the systematic discrepancy between
    the X-ray luminosities of  accreting and non-accreting young
    stars. Unlike the previous EUV models (e.g. Alexander et al. 2006b;
    Alexander \& Armitage, 2009), the X-ray photoevaporation model can
    account for a large fraction of the observed `transition'
    discs. In particular,
photoevaporation can account for the population of small inner holes
$R<20$AU, 
those with low accretion rates $\dot{M}_*<10^{-8}$\msunyr and those
with gas inside the inner-hole. However, `transition' discs with large
inner holes and large accretion rates are likely to be created through
a different mechanism (e.g. Espaillat et al. 2010).     

{ The first direct evidence and best test of individual
    photoevaporation models is the detection of a blue-shifted
    12.8$\mu$m NeII line from
the surface of the disc around TW Hya (Pascucci \& Sterzik 2009), which is
consistent with either  an EUV or X-ray driven photoevaporative wind
(Pascucci et al. 2011)}. Moreover, Ercolano 
 \& Owen (2010) recently showed that  the 
previously unexplained low velocity component of the OI 6300\AA, - 
emission  blue-shifted by $\sim5$km s$^{-1}$ detected by Hartigan et al. 1995
around most primordial discs - is naturally produced in an X-ray driven
photoevaporative flow. The observed luminosity is too high to be
consistent with an EUV wind and the blue-shift is too large to be
consistent with an FUV wind. { Currently, the only well resolved OI
    6300\AA~ line is detected from the accreting `transition' disc
    around TW Hya, where the line presents with no blue-shift (Pascucci et
    al. 2011).  Unfortunately, hydrodynamic models of the inner regions of photoevaporating
   transition discs during
    the accreting phase do not exist for any of the photoevaporation models (only transition discs with cleared inner gas discs were considered by Ercolano \& Owen 2010) and useful
    comparisons must wait until these calculations are completed.
    } Therefore, pending these calculations, and the availability of high resolution spectroscopic observations of many objects, direct comparisons of
predicted to observed OI line profiles are not possible.   

While observations are providing a fruitful avenue for direct
comparisons of photoevaporation models, in this work we develop a
theoretical framework in which we can analyse the combined effects off
all three heating mechanisms (EUV,FUV and X-rays). The most relevant
scale in any photoevaporation model is the radius at which a gas of a
given temperature ($T_{\textrm{gas}}$) becomes unbound from the
central star and is given by\footnote{Throughout this work we use $\{R,\phi,z\}$ to represent the cylindrical co-ordinate system and $\{r,\theta,\varphi\}$ to represent the spherical co-ordinate system, in all cases centred on the star.}: 
\begin{equation}
r_g=8.9 \textrm{AU} \left(\frac{T_\textrm{gas}}{10^{4} \textrm{ K}}\right)^{-1}\left(\frac{M_*}{1\textrm{ \msun}}\right)
\end{equation}
For the EUV photoevaporation model this becomes a fixed radius (for a given disc), since
    EUV heating gives rise to essentially isothermal gas at
    $10^4$K. For the X-ray and FUV model it is better to think of this
    radius as an escape temperature scaled length, allowing the
    consideration of different radial scales around different mass
    stars. However, in order to be consistent with the previous EUV
    only literature { (Hollenbach et al. 1994; Clarke et al. 2001; Font
    et al. 2004; Alexander et al. 2006a)}, we separately fix $R_g=8.9(1/1\textrm{\msun})$AU. 
Furthermore, as we will
    discuss in detail  in Section 2, this radius has an important physical
    significance as it is a measure of the  radius at which a flow at
    temperature $T_{\textrm{gas}}$ can go through the sonic
    surface. For X-ray heating in the range
    $T_{\textrm{gas}}\sim1000-10000$K, $r_g\sim 1-10$AU for a very low
    mass star (0.1\msun) and $r_g\sim10-100$AU for a solar type
    star. For FUV heating,  $T_\textrm{gas}\apprle 1000$K,
    $r_g\apprge10$AU for a very low mass star and $r_g\apprge100$AU
    for solar type stars. 

In this work,  we lay out the theoretical foundation for  
photoevaporation, deriving  
the scaling with X-ray
    luminosity and stellar mass
in Sections~2. In Section~3 we discuss the likely interplay
between the effects of combined 
    EUV, FUV and X-ray heating on the calculation
    of photoevaporation rates. Section~4 presents the results of
    radiation-hydrodynamic calculations that test the theoretical
foundations laid out in the  Section 2, while Section ~5 contains
some simple numerical experiments that address the discussion of
combined FUV and X-ray heating contained in Section 3.  Section~6 describes the final clearing of the outer disc by `thermal sweeping'.
In Section~7 we describe the implications of photoevaporation  on disc evolution and dispersal. In Section~8 we discuss the presented results and draw our conclusions in Section~9.

In Appendix~A we discuss the possibility of self-similarity in X-ray heated flows and in Appendix~B we provide numerical fits to our photoevaporation calculations for use by the community.

\section {The theory of X-ray photoevaporation}

The problem of thermally driven winds from discs has been studied analytically in
    the past (Begelman et al. 1983; Liffman 2003; Adams et
    al. 2004). While a full analytic solution to the problem has not
    been obtained in even the most simplistic thermodynamic
    constructions, considerable progress can still be made. In the
    Appendix we discuss the existence of similarity solutions with
    regard to the problem of X-ray photoevaporation. The results
    from the analysis are as follows:
\begin{itemize}
\item For primordial discs, the total mass-loss rates scale linearly with X-ray luminosity
    and are independent of stellar mass. 
  \item For primordial discs, the density at a given value of $\mathbf{r/R_g}$ scales
    with X-ray luminosity and stellar mass as $n(L_X,M_*)\propto
    L_X\,M_*^{-2}$ for primordial discs. 

\item For discs with inner holes of fixed radius, the total mass-loss scales linearly with the X-ray luminosity.

\item Discs with inner holes with a range of radii are not strictly mutually self-similar: see section 4.3.1 for an approximate self-similarity argument that can be used to understand the numerical results in this case.

\end{itemize}

In this section, we derive the scalings listed above using a series of simple analytic arguments. 
 Since the mass-flux is conserved in any steady-state photoevaporative
    flow we are free to evaluate this quantity at any point in the
    flow. Therefore, choosing to do this at the sonic surface the
    total mass-loss rate is given by:

    \begin{equation}
      \dot{M}_w=\int_Sn_s(\mathbf{r_s})c_s(\mathbf{r_s})\mathbf{l}(\mathbf{r_s})\cdot\dd\mathbf{S}
\end{equation}
where $S$ is the sonic surface, $n_s$ \& $c_s$ are the density and
    sound speed at the sonic surface respectfully, and $\mathbf{l}$ is
    the unit vector along a streamline. In the case of a flow
    from a disc, this can be approximately written as:
\begin{equation}
\dot{M}_w=2\int_0^\infty 2\pi R n_s(R)c_s(R)\dd R\label{eqn:m_tot}
\end{equation} 
where we note that the true mass loss flux will differ from this by a factor
of order unity due to the fact that the flow is not purely vertical at the
sonic surface. Therefore, provided we can determine the flow
    properties at the sonic surface we can estimate the total
    mass-loss rates.

\subsection{The Sonic Surface in X-ray Heated Flows}
The sonic point is a critical point of the steady-state momentum
    equation when $u=c_s$, and physically arises as a consequence of
    wave-steeping requirements, namely that no strong shock forms
    (Landau \& Lifshitz, 1987).

Several authors have derived the conditions on the sound speed at the sonic surface for
    a flow emerging from a disc (e.g. Begelman et al. 1983; Adams et al. 2004; also
    shown in Appendix A for completeness). Under the conditions expected for disc
    winds (sub-sonic launching, flow time-scale $>$ thermal
    time-scale, height of the sonic surface ($z_s$) $\ngg\,\, R$) the sound speed at the sonic surface should be to
    order unity the Parker value at each cylindrical
    radius, namely:

\begin{equation}
c_{s}^2 \approx {{GM_*}\over{2R}}\label{eqn:cs}
\end{equation}
Therefore, we can {\it a priori} determine the sound speed, and hence
    the temperature at the sonic surface.  

\subsubsection{Density-Sound-Speed Coupling in X-ray Heated Flows}
If we can specify both the
    density and sound speed at the sonic surface then we have
    determined the mass-flux in the entire flow as mass-flux is
    conserved. We argued above that in the case of disc winds, the sound speed at the sonic surface is roughly given by the
    Parker value (Equation~\ref{eqn:cs}). This means that the gas temperature at
    the sonic surface is also simply given as a function of
    cylindrical radii.

\begin{figure}
\centering
\includegraphics[width=\columnwidth]{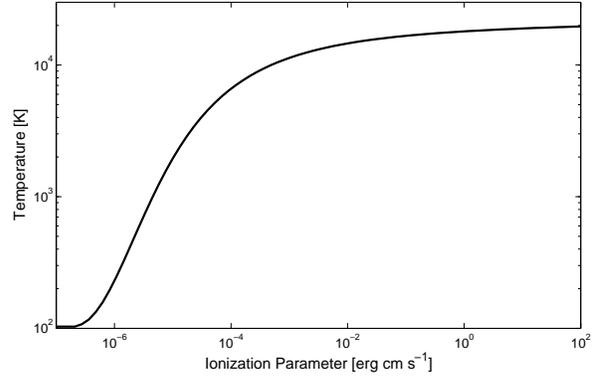}
\caption{Fit to temperature-ionization parameter calculated by Owen et
    al. (2010).}\label{fig:ionT}
\end{figure}

It is well known that the temperature of X-ray heated gas can be
    described in terms of the local ionization parameter (e.g. Tarter
    et al. 1969), given by:
\begin{equation}
\xi=\frac{4\pi F_X}{n}
\end{equation}
where $F_X$ is the X-ray flux ($L_X/4\pi r^2$). Ercolano et al. (2009ab) \& Owen et
    al. (2010) found that the temperature of gas within a column
    density to the central star of $10^{22}$ cm$^{-2}$ (this being the penetration depth of 1~KeV photons\footnote{see discussion in Section 4.2.2 for implications of attenuation which is neglected here}) to the
    central star can be described in terms of optically thin X-ray heating. Thus, adopting the assumption of optically thin X-ray heating  the ionization
    parameter becomes $\xi=L_X/n r^2$, and the gas temperature is
    specified in terms of local variables only, where the form of this
    profile $T=f(\xi)$ is shown in Figure~\ref{fig:ionT}. 

 At the
    sonic surface (provided $z_s\ngg R$ so $R\sim r$) we may write:
\begin{equation}
c_s^2=\frac{k_Bf\left(L_X/n_s R^2\right)}{\mu
    m_h}=\frac{GM_*}{2R}\label{eqn:den1}
\end{equation}
Therefore, we may re-arrange Equation~\ref{eqn:den1} to give an
    expression for the density at the sonic surface\footnote{where
    $f^{-1}\circ f =1$ and $f^{-1}\equiv \hspace{-0.28cm} /\hspace{0.2cm} 1/f$}:
\begin{equation}
n_s=\frac{L_X}{R^2}\left[f^{-1}\left(\frac{GM_*\mu
    m_h}{2k_BR}\right)\right]^{-1}\label{eqn:den_s}
\end{equation}
One may also write this at fixed $R/R_g$ to obtain the mass-scaling
    of the density, namely $n\propto M_*^{-2}$. Thus, provided we know the form of $f(\xi)$, the density at the sonic
    surface is also fixed {\it a priori} through the
    temperature-ionization parameter relation, which provides a density-sound-speed coupling.
\subsubsection{Consequences of Density-Sound-Speed Coupling}
The density-sound-speed coupling found at the sonic surface in
    optically thin  heated X-ray
    heated flows provides some very powerful constraints on the flow
    properties of X-ray photoevaporative winds, and implies that the
structure and temperature of the underlying disc has little effect
on the resulting flow.

    This arises since in any steady state flow mass-flux must be
    conserved; therefore, in this case since we are able to described the
    mass-flux at the sonic surface {\it a priori} with reference to
    the X-ray physics alone, we have  determined the mass-flux in the
    {\it entire} flow without reference to the disc properties.

This picture is very different to pure EUV (Hollenbach et al.  1994) or FUV
    (Gorti \& Hollenbach 2008,2009) flows where the underlying density 
    structure in the launch region is crucial to the mass-loading of the flow. Unlike
    EUV/FUV flows where the flow will adjust to changes in the
    underlying disc's structure, an X-ray flow will adjust the
    underlying disc structure to `feed' the flow at the required
    rate, { provided the sonic surface remains optically thin to the X-rays responsible for the heating}. 

\subsection{Estimates of the total mass-loss rate}
Armed with the knowledge that it is the X-ray physics, and not the
    disc structure that sets the overall mass-loss rates we are free
    to integrate Equation~\ref{eqn:m_tot} over the sonic surface to
    estimate the total mass-loss rates. We can use
    Equations~\ref{eqn:cs} \& \ref{eqn:den_s} to substitute for $n_s$ and $c_s$
    in the integrand. Furthermore, if we change the
    variable of integration in Equation~\ref{eqn:m_tot} from
    cylindrical radius $R$ to escape temperature we find: 
\begin{eqnarray}
\dot{M}_w(M_*,L_X)\!\!\!\!\!&=&\!\!\!\!L_X\sqrt{\frac{8k_B\pi^2}{\mu m_h}}\int_0^\infty\!\!\!\dd\Tesc\,
   \frac{\Tesc^{-1/2}}{ f^{-1}(\Tesc/2)}\label{eqn:inttesc}\\
&\approx&8\times 10^{-9}\left(\frac{L_X}{1\times 10^{30}\,\textrm{erg
    s}^{-1}}\right) \textrm{M}_\odot\,\textrm{yr}^{-1}\label{eqn:tot}
\end{eqnarray} 
(where the second result comes from evaluating the integral out to an
    escape temperature of 1000K, or a radius of $\OO (100)$AU for a
    solar mass star). This mass-loss rate  is in good agreement with the numerical
    result of $6.7\times 10^{-9}$M$_\odot\,$yr$^{-1}$ obtained from
the hydrodynamic simulations of Owen et al. (2011b) and is also consistent
with the approximately linear scaling between mass-loss rate and X-ray
    luminosity found in the simulations. Furthermore, the results
    indicates there is no explicit dependence between stellar mass and
    mass-loss rate (although the $M_*-L_X$ scaling
    introduces a strong positive implicit scaling). We also derive
    this scalings in more detail through the existence of similarity solutions  presented\footnote{
    Note that the scalings presented in Appendix~A do
not rely on the approximate expression for the temperature at
the sonic surface contained in Equation\ref{eqn:cs}.} in Appendix A.

\subsection{Importance of Numerical Results}\label{sec:Theory}

   In Sections 4 and 5 we present hydrodynamical simulations that
have been designed to test and refine the ideas set out above. However,
the role of the numerical simulations is not merely one of  validation:
although Equation~\ref{eqn:tot} is of great utility in predicting the integrated
mass-loss from the system (and can be used to derive the mass loss
rate per unit area {\it at the sonic surface}), it does not  - without
knowledge of the detailed streamline topology below the sonic surface -
allow one to calculate the mass loss rate per unit area from the disc
itself ($\dot \Sigma_w(R)$). { For example the form of the
    integrand in Equation~\ref{eqn:tot} suggests that the mass-loss is
    dominated in the lower temperature regime (2000-5000K), but these
    stream-lines connect the sonic surface to the disc surface in
    { a non-uniform way, such that there is not a one to one correspondence between the mass-flux profile at the sonic surface and the mass-flux profile from the disc.} This results in a mass-loss profile that peaks at
    smaller radii in the disc, rather than the large radius suggested
    by the calculations at the sonic surface.} Although this may appear to be a detail, it turns out that,
when photoevaporation is combined with secular disc evolution,
the pattern of disc clearing is rather sensitive to the form of
$\dot \Sigma_w(R)$. { In particular, it is the form of $\dot{\Sigma}_w(R)$ that determines } where the gap first
opens in the disc, the extent to which the system undergoes a period
of `photoevaporation-starved accretion' prior to gap opening and hence
the surface density profile of the disc outside the gap. This latter then
determines the rate at  which the disc is evaporated once the inner
disc is completely cleared out. This is of obvious importance if one
wants to compare the predictions of photoevaporation theory with the
observed incidence of transition discs with holes of different
sizes (see Section 8.1).

\section{Theory of UV \& X-ray Photoevaporation}\label{sec:TheoryXUV}
We now turn to considering the situation where X-ray heating is combined
with ultraviolet heating, considering in turn the case the ionising
(EUV) and non-ionising (FUV) continua.

\subsection{EUV and X-ray heated flows}

  The { radiative transfer calculations} of 
    Ercolano \& Owen (2010)
    demonstrated  why, in the presence of an X-ray flow, EUV
    heating can be neglected. 
\begin{figure}
\centering
\includegraphics[width=\columnwidth]{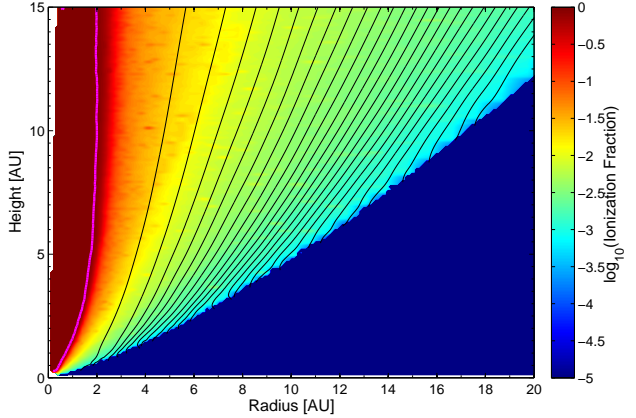}
\caption{Ionization structure of the X-EUV wind around a 0.7\msun star with an X-ray luminosity of $2\times10^{30}$erg s$^{-1}$, the streamlines are equispaced in 2\% of the cumulative mass-loss rate and the magenta contour shows the region of 98\% ionization fraction (i.e. roughly the penetration depth of EUV photons). This shows that an X-ray driven wind is itself entirely optically thick to EUV photons and prevents them from reaching the disc surface}\label{fig:ioniz}
\end{figure}
 Figure~\ref{fig:ioniz} shows the
    ionization structure of an X-ray wind subject to combined X-ray and EUV irradiation,
with the magenta contour indicating the point at which the flow
develops a significant neutral fraction and beyond which the EUV cannot
penetrate. { This is easily understand if we consider the Str\"omgren
    radius:}
 \begin{equation}
   r_s\approx 0.2\,\,{\rm AU}\left(\frac{\Phi_*}{10^{41}\,{\rm
    s}^{-1}}\right)^{1/3}\left(\frac{n}{10^8\, {\rm
    cm}^{-3}}\right)^{-2/3}
\end{equation}
{ where $n=10^8$ cm$^{-3}$ is the density at the base of the flow
    for an X-ray luminosity of $L_X=2\times10^{30}$erg s$^{-1}$. Now, Equation~\ref{eqn:den_s} (or Appendix~A) implies the density
    scales linearly with X-ray luminosity ($n\propto L_X$). If we furthermore assume that the EUV and X-ray luminosity are both coronally produced
    (e.g. Alexander et al. 2005), so the luminosities scale linearly we find
    $r_s\propto L_X^{-1/3}$. This means that even at the lowest X-ray
    luminosities ($\sim10^{28}$~erg s$^{-1}$) the EUV penetration is
    still too small to penetrate the X-ray flow. Even if we adopt the strongest EUV and the weakest X-ray luminosities quoted in the literature (i.e. $L_X=10^{28}$ erg s$^{-1}$, $\Phi_*=^{10^44}$ s$^{-1}$; G\"udel et al. 2007; Alexander et al. 2005) we find that 
    the EUV flux cannot reach the surface of the disc from which the bulk of the X-ray wind is launched. { Therefore, EUV irradiation can be neglected when X-ray radiation is also present. }}

\subsection{FUV and X-ray heated flows}

\begin{figure}
\centering
\subfigure[X-ray dominated photoevaporation]{\includegraphics[width=\columnwidth]{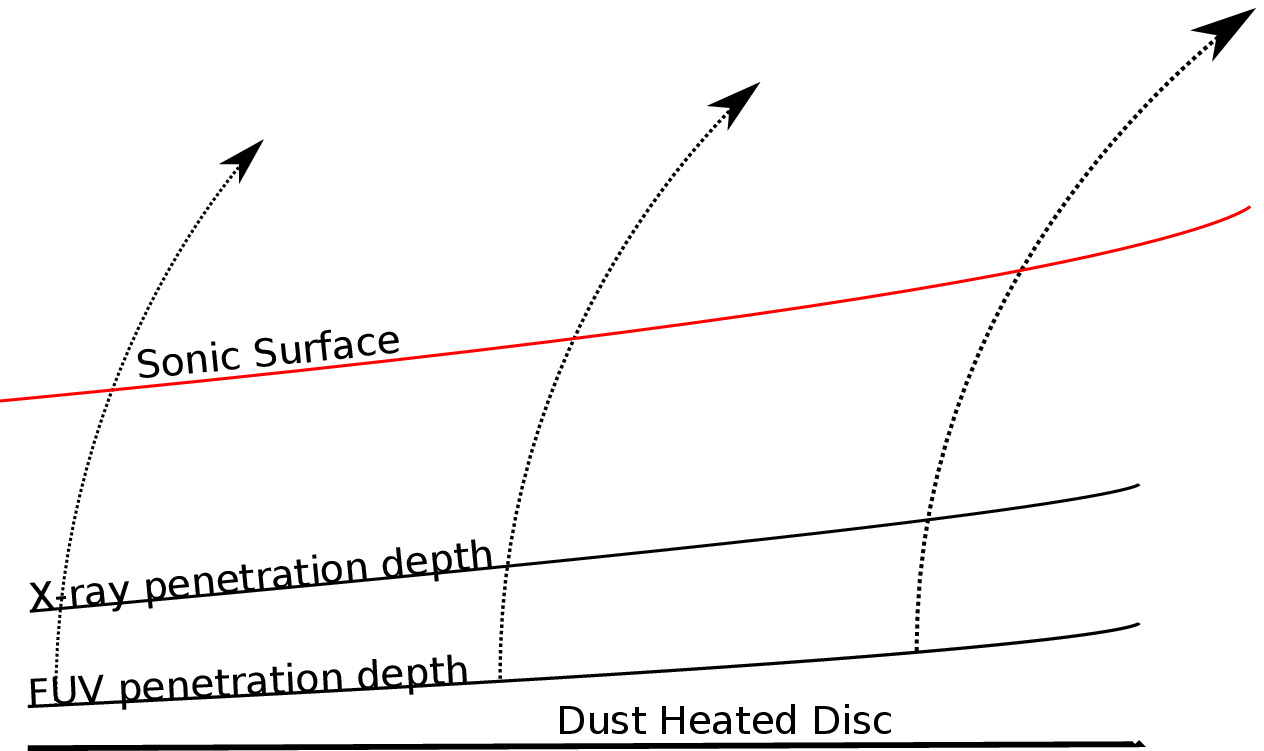}\label{fig:X-rayfuva}}
\subfigure[FUV dominated photoevaporation]{\includegraphics[width=\columnwidth]{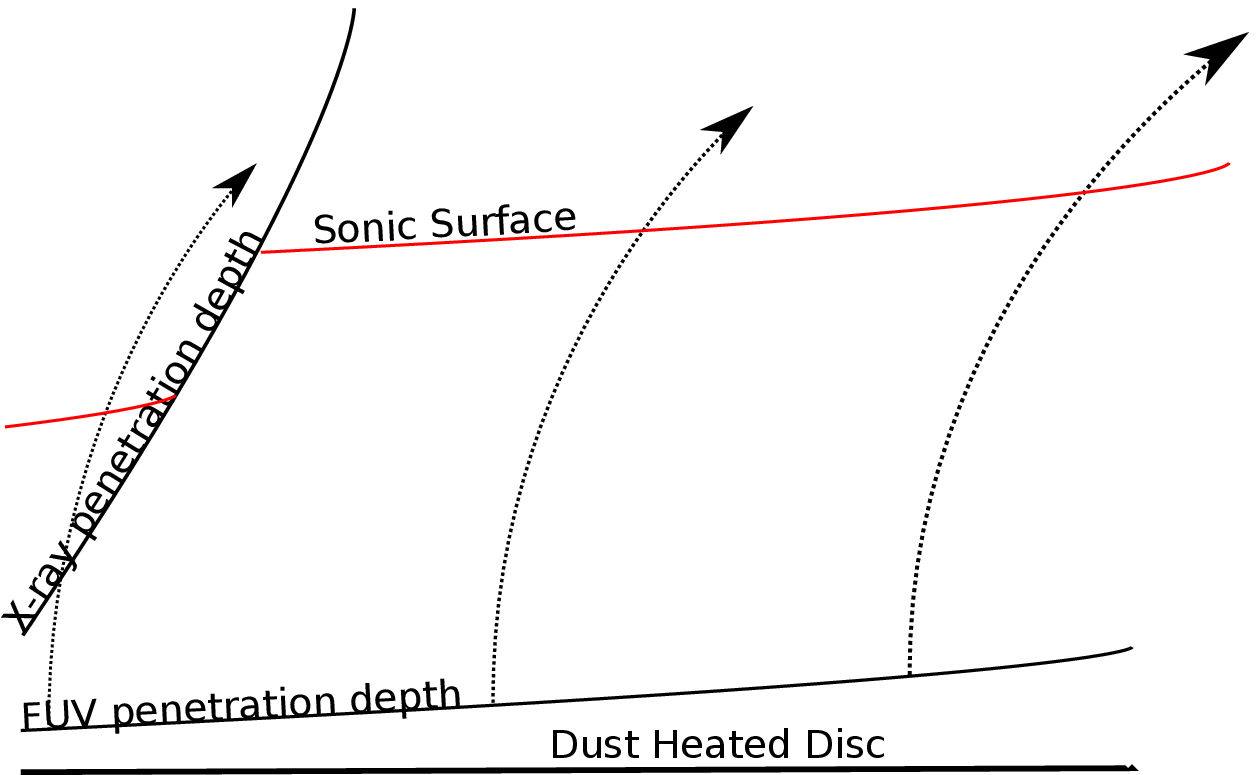}\label{fig:X-rayfuvb}}
\caption{Figure showing a cartoon sketch of photoevaporation from a disc exposed to UV and X-ray irradiation. The flow begins in the FUV heated region, but at some height the flow transitions to the X-ray heated region and then passes through the sonic surface in the X-ray heated region.}\label{fig:primcartoon}
\end{figure}
 Gorti \& Hollenbach (2008,2009) considered the combined effect of UV and
X-ray heating in setting the hydrostatic disc structure and then used
their static structures to predict resulting flow rates. { In the case of FUV heating, the underlying heated disc structure determines the properties at the sonic surface and a variety of solutions, whether static
or flowing, asymptote to the same (nearly hydrostatic) structure
at small $z$. }Although the results of Gorti \& Hollenbach still represent
the state of the art in terms of the thermal and chemical
structure of X-ray and FUV irradiated discs, the implications of these studies
for the resulting mass loss rate are  currently unclear. In Section~5
we explore this further by using a simple parametrisation of FUV
heating in our radiation hydrodynamical simulations {(although we caution this should not be regarded as a proper treatment of the FUV heated flow)}. Here
we discuss the general theoretical framework for combined
FUV and X-ray heated flows.

  X-ray heating is effective up to a total hydrogen column
of around $10^{22}$ cm$^{-2}$. For models using depleted dust in the
disc atmosphere (as in the models of D'Alessio et al. 1998,1999,2001), FUV heating extends to significantly higher columns. 
This means that travelling along a line from the star to a point
in the disc  one first traverses gas that is exposed to both X-rays\footnote{By X-rays we are referring to the X-ray photons that provide significant thermal heating with a maximum energy of 1-2KeV and not those at higher energies that can penetrate larger columns but provide insignificant heating (see Ercolano et al.  2009b).} and
FUV, then to FUV only and then to neither. 
Within  
$\sim 100$ AU  the X-rays dominate the heating in the regions that are exposed to
both X-rays and FUV. Therefore, in this region the sequence (as one
proceeds away from the star) is X-ray heated, FUV heated and then
heated by neither - in reality heated by coupling to the dust - (e.g. Gorti \& Hollenbach, 2009). 

 Though we expect this sequence of X-ray heating followed by FUV
heating, there are still two
topologically distinct possibilities which we illustrate\footnote{Note the similarity of this discussion
to the analysis of flow from proplyds in the case of combined
irradiation by FUV and EUV radiation fields: see Johnstone et al 1998}
in Figure 3.  
Figure~\ref{fig:X-rayfuva} is a schematic illustration of the case where the flow in
the FUV heated region never exceeds its local sound speed for any
of the streamlines shown. The sonic transition occurs in the X-ray heated
region. In this case we can use the previous arguments to find
that the sound speed at the X-ray heated sonic surface is simply given
by Equation~\ref{eqn:cs} to order unity as before: in other words the temperature
at the sonic surface is simply a function of cylindrical radius. Since
this also fixes the density at the sonic surface this implies that the mass flux is just a function of $R$ and
X-ray luminosity. In other words, whatever the detailed physics of the
heating and cooling processes in the X-ray dark region, they must combine
to drive a flow that is able to deliver the required mass flux to the
X-ray sonic surface. We therefore expect that if the geometry is as shown
in Figure~\ref{fig:X-rayfuva}, the photoevaporation rate is entirely controlled by
X-ray heating. 
  
 On the other hand, Figure~\ref{fig:X-rayfuvb} shows three streamlines where the
two at larger radius undergo a sonic transition within the FUV heated
region. 
In this case the flow has to self-adjust to deliver the required
conditions at the FUV sonic surface - in other words it is the structure
of the FUV heated region that controls the mass flow rate. Although
in principle gas flowing out in this FUV heated wind could become X-ray
heated at some point, this region would not be causally connected
to the flow base, the two regions being separated by a shock.  However, it is unlikely that an FUV driven
wind would ever be significantly heated by X-rays from the central star.
This is because we require the condition at the sonic surface in the
FUV heated region to satisfy Equation~\ref{eqn:cs}. Given that FUV heating does not
generally attain temperatures in excess of $\sim 1000-2000$K (Gorti \&
    Hollenbach 2004; Bruderer et al. 2009), this implies that the
schematic FUV sonic surface shown in Figure~\ref{fig:X-rayfuvb} must lie at radii of
$\apprge100$ AU where X-rays are incapable of heating the gas to greater
temperatures than the FUV.

  We therefore conclude that FUV driven winds are only likely to  be significant at
radii $> 100$ AU.   If the flow topology at smaller radii
is as shown in Figure~\ref{fig:X-rayfuva} then the photoevaporation rate is set entirely
by the X-rays. The only situation in which FUV heating could control
mass loss from much smaller radii is if the streamlines crossing the
FUV sonic surface at $> 100$ AU actually originated at small radii
Naturally, we need to perform a radiation-hydrodynamic simulation
with FUV heating to assess this possibility (Section~6).

\section{Numerical Models of Photoevaporating Discs}
As discussed in Section~\ref{sec:Theory}, numerical radiation-hydrodynamic simulations are required  to obtain accurate mass-loss rates and importantly the mass-loss profile. In this section we extend the parameter space investigations by Owen et al. (2010,2011b) to cover the entire range of X-ray luminosity, mass \& inner hole size expected for discs around low-mass stars, and thus test and calibrate the theory discussed above. Owen et al. (2011b) already investigated the explicit dependence of mass-loss rate on X-ray luminosity and inner-hole size around a 0.7M$_\odot$ star. Therefore, in this work we are only concerned with extending the models to lower mass and varying disc structures, and we have performed simulations of photoevaporating discs around 0.1 \& 0.7 M$_\odot$ star.  

\subsection{Numerical Method}
The numerical method is similar to that described in Owen et al. (2010), and is briefly described here for completeness. The simulations were performed using a modified version of the {\sc zeus-mp/2} code\footnote{Note: In order to increase computational efficiency we have moved from the scalar {\sc zeus-2d} code to the MPI {\sc zeus-mp/2}, which parallelises efficiently on up-to 64 processors.} (Hayes et al. 2006), where the modifications are described in detail in Owen et al. (2010). X-ray heating is included using a temperature-ionization parameter relation up-to a critical column of $N=10^{22}$cm$^{-2}$; at columns larger than this value the gas temperature is fixed to the dust temperature. We start from the hydrostatic disc calculations of D'alessio et al. (2001) that only include dust heating; this disc structure is then irradiated with X-rays which drive a photoevaporative wind that evolves to steady state. We also use spherical co-ordinates in order to match the symmetry of the solution at large radius.
\subsubsection{Primordial Discs}
In order to test the mass-scaling, along with the predicted X-ray
luminosity scaling in different mass regimes, we perform 5
simulations of photoevaporation for a primordial disc around a
0.1M$_\odot$ star, evenly spaced in logarithmic X-ray luminosity from
$\log_{10}{L_X}$ = 28.3 to 30.3, which covers the range of observed X-ray luminosities around M-type stars. We use a grid with 288 logarithmically spaced cells in the radial direction and 144 cell equally spaced in the angular direction. The grid extends from $r=0.03$AU ($\sim0.03R_g$) to $r=46$AU ($\apprge45R_g$) and thus covers the range where we expect photoevaporation to be important. 

In order to consider the effect of varying the disc structure we run three further simulations of the original calculation performed in Owen et al. (2010), i.e. a disc around a 0.7\msun star with an X-ray luminosity of $2\times10^{30}$erg s$^{-1}$: firstly we increase and decrease the dust temperature by a factor of $\sqrt{2}$, since the dust in the region that interacts with the X-ray heated region is optically thin. These changes in temperature effectively correspond to a change in the stellar luminosity by a factor of 4. We also consider a simulation with a disc mass of 10\% of the original calculation i.e. $M_d=2.6\times10^{-3}$\msun. 
\begin{figure}
\centering
\includegraphics[width=\columnwidth]{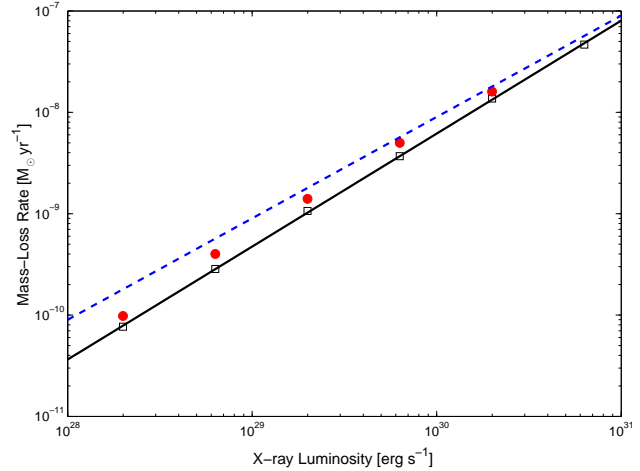}
\caption{Mass-loss rate plotted against X-ray luminosity, the open squares show the simulations results for a 0.7\msun star obtained in Owen et al. (2011b), the filled circles show the simulation results for a 0.1\msun star described in this work. The solid line shows the fit to the mass-loss rates used for a population-synthesis study conducted in Owen et al. (2011b), the dashed line shows the order of magnitude estimate obtained in Section~3.}\label{fig:mdot_tot}
\end{figure}
\subsubsection{Discs with inner holes}
 Since the low stellar mass simulations presented here have lower
disc masses than the higher mass counterparts described in Owen et al (2011b),
we have to take into account of the fact that the density structure of
the underlying disc may be being eroded during the hydrodynamical
simulation on a time-scale less than the time required to set up
a steady state wind (this latter being $\sim$ the sound crossing
time to the edge of the grid; see discussion in Alexander et al 2006a). 
Such erosion is not necessarily `realistic'
since we do not model the viscous effects in the disc which may re-supply
material to the disc inner edge.  In order to explore this effect we consider
three classes of simulation: one with a small inner hole ($0.7 $ AU)
where the disc is not significantly depleted over the duration of the
simulation, one starting with inner hole radius of $3$ AU that was
allowed to erode over the duration of the simulation and one with
radius of $5$ AU where the inner edge is `held' at $5$ AU by artificial
re-supply (i.e by re-setting the X-ray dark `base density' to its
original value every time-step) to obtain a steady-state simulation. For each of these cases we model
two X-ray luminosities ($\log L_X = 29.3,30.3$).
\begin{figure*}
\centering
\includegraphics[width=\textwidth]{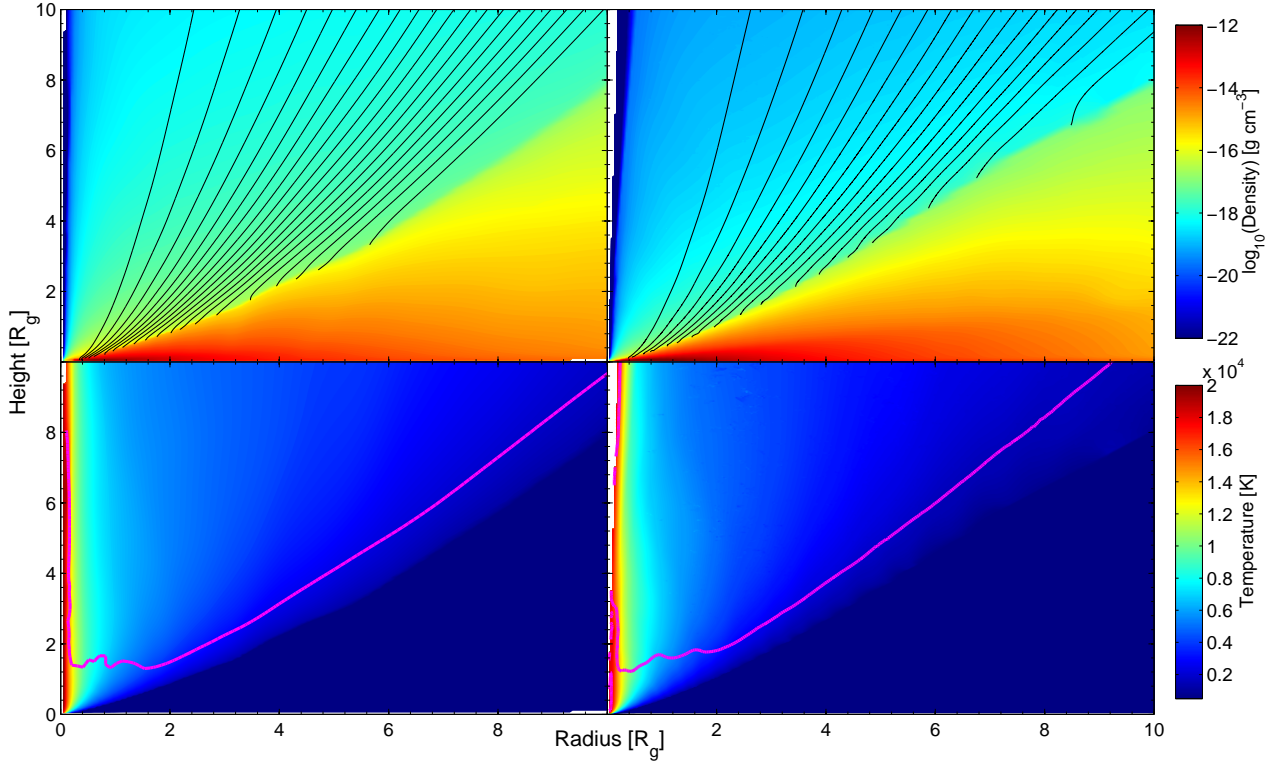}
\caption{Figure showing the flow topology around a 0.1 (left panels)
    and 0.7 (right panels) \msun star for X-ray luminosity of
    $2\times10^{29}$erg s$^{-1}$ and $2\times10^{30}$erg s$^{-1}$
    respectively; the results for the 0.7\msun star are taken from Owen et al. (2011b). The top panels show the density structure with streamlines at 5\% intervals of the cumulative mass-loss rate. The lower panels show the temperature structure of the flow, where the solid line represents the sonic surface. We note  (in agreement with the theoretical model presented in Section ~2) that 
 the mass-loss profile and temperature are independent of stellar mass when considered in terms of a radius scaled by $r_g$, but the density structure scales as $M_*^{-2}$.}\label{fig:compare}
\end{figure*}

\subsection{Simulation results: primordial discs}
In Figure~\ref{fig:mdot_tot} we show how the mass-loss rate scales with both mass and X-ray luminosity; the results confirm the theoretical predictions described in Section~2.2 to first order, namely no explicit mass-dependence on the photoevaporation rate and furthermore a mass-loss rate that scales approximately linearly with X-ray luminosity.

In Figure~\ref{fig:compare} we compare the flow morphology for an
    X-ray luminosity of $2\times10^{29}$erg s$^{-1}$ around the 0.1
    (left panels) and 0.7\msun (right panels) stars. The possible existence of
    self-similarity described in Appendix~\ref{sec:TheoryX} implies that the temperature of the flow should be fixed by the value of the effective potential, thus to first order the temperature in the flow should be fixed by the mass of the central star and thus be scale free when considered in terms of a radius scaled by $r_g$. The lower panels of Figure~\ref{fig:compare} show that to first order this the case. Furthermore the top panels show the density structure in the flow and the streamlines at $5 \%$ intervals of the cumulative mass loss
rate. 
{ Despite small differences the similarity of the right and left hand panels demonstrate that the simulations are close to being self-similar. Moreover, the density is 
lower at the higher mass (at given $R/R_g$) by a factor consistent with
the $M_*^{-2}$ scaling predicted in Section ~2 and Appendix~A. We thus find that the simulations
are in good agreement with the self-similarity arguments presented in
    Appendix~A.}

\subsubsection{Dependence on disc structure}

As first indicated by the numerical models calculated in Owen et al. (2010) with adiabatic discs, the disc structure has very little bearing on the 
final mass-loss rate. {  In Section~2 we developed an argument
to explain this insensitivity to underlying disc structure, invoking
the necessity to satisfy particular conditions on the X-ray sonic
    surface.}
Here we test this idea further by exploring the winds produced by more realistic variations in disc structure.

\begin{figure}
\centering
\includegraphics[width=\columnwidth]{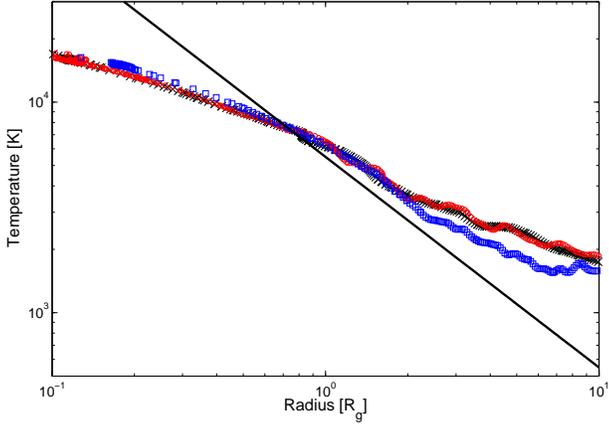}
\caption{The temperature at the sonic surface as a function of cylindrical radii (scaled by $R_g$) for 3 models with two different stellar masses (0.1 \& 0.7\msun) and X-ray luminosities of ($2\times10^{29}$ \& $2\times10^{30}$ erg s$^{-1}$). The open squares show the original 0.7\msun, $2\times10^{30}$erg s$^{-1}$ calculated by Owen et al. (2010), the black crosses are those of the 0.1\msun $2\times10^{29}$erg s$^{-1}$ and the open circles are for the  0.7\msun, $2\times10^{30}$erg s$^{-1}$ model calculated with at disc depleted in mass by a factor of 10 from the original calculation. The black line shows the temperature expected from the order of magnitude calculations performed in Section~2. We note all primordial disc calculations have sonic temperatures that follow similar loci.}\label{fig:sonic_srf}
\end{figure}

These correspond to variations in the stellar luminosity and
    underlying disc mass. Given the theoretical description of the
    flow predicts that the sonic surface controls the mass-flux, in
    Figure~\ref{fig:sonic_srf} we show the temperature at the sonic
    surface as a function of cylindrical radius. The black
line
shows the escape temperature as a function of radius. We see that
beyond $\sim R_g$ the models are close to the escape temperature, in line
with the expectation of Equation~4. At smaller radii, the escape
temperature rises more steeply with decreasing radius than the models.
We can simply understand this change in behaviour by noting that
it occurs at a temperature of around $10^4$K, which is where the ionisation
parameter temperature relation flattens (see Figure~1). At higher
temperatures
the temperature is rather insensitive to ionisation parameter. Thus,
whereas at larger radius, the ionisation parameter adjusts to produce
the temperature at the sonic surface demanded by Equation ~4, at small
radius the temperature is more or less fixed. In this case, the Parker
wind sonic transition corresponds to an almost constant spherical
radius. This transition can be seen in the topology of the
sonic surface at $R \sim R_g$ in Figures~5 \& 8. We further demonstrate
this behaviour in Figure~7 which plots, for a variety of models, the
effective potential on the sonic surface as a function of cylindrical
radius - evidently all models evolve so as to satisfy a common relation
on the sonic surface, regardless of the mass and temperature of the
underlying disc.

  Figure~8 illustrates the flow topology for the models with various
disc temperatures and disc mass. Evidently these are qualitatively similar
but differ in detail.  For example, the sonic surface is at lower height
and the density at intermediate height above the disc plane is reduced
when the disc mass is lowered. This however has a minimal effect on the
disc mass loss profiles: even a change in height of the sonic surface by
$20 -30 \%$ has a small effect on the effective potential on the sonic
surface and hence hardly affects the associated temperature, density and
mass
flux.  
\begin{figure}
\centering
\includegraphics[width=\columnwidth]{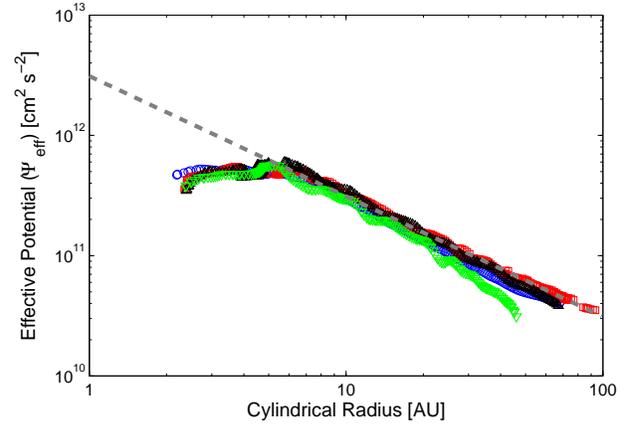}
\caption{The value of the effective potential ($h^2/2R^2-GM_*/r$ - note it is shown as a positive value for simplicity) at the sonic surface
    plotted against cylindrical radius for various underlying disc
    structures. The blue points are for the original calculation of
    Owen et al. 2010, the black points represents the mass depleted
    disc described above, the red points show the disc with increased
    dust temperature while the green points show the disc with
    decreased dust temperature as described above. The dashed line
    shows the prediction from the model described in Section~2. }\label{fig:sonic_eff}
\end{figure}
Figure~\ref{fig:sonic_eff} shows that for most radii the value of the effective potential is
    as expected. The dip at $R< 5 AU$ represents the region described
    above that is unable to reach the required escape temperature and
    thus the sonic surface occurs at a constant height and the $z\ngg R$
    assumption breaks down.
\begin{figure*}
\centering
\includegraphics[width=\textwidth]{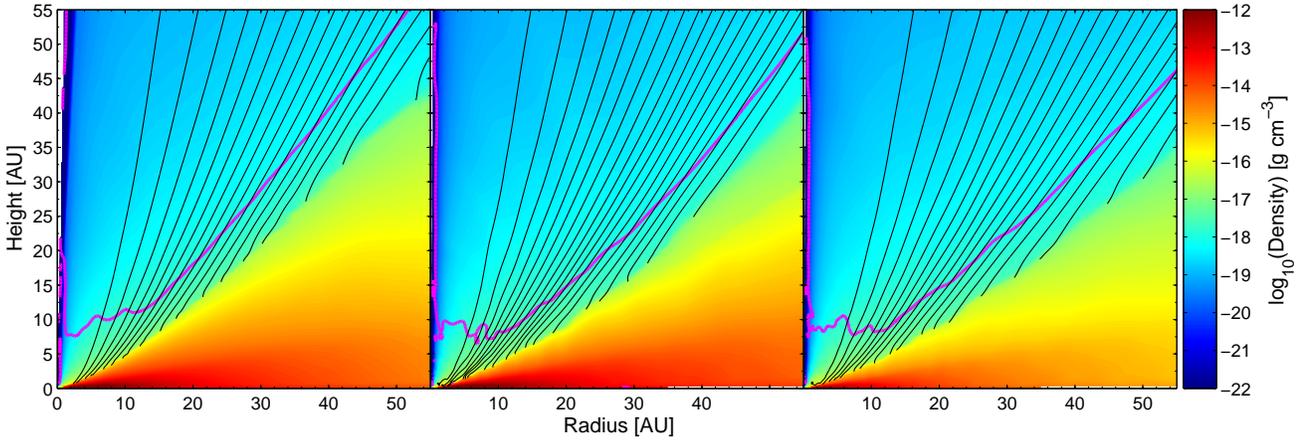}
\caption{Density and velocity structure for photoevaporative winds with same stellar parameters ($M_*=0.7$\msun, $L_X=2\times10^{30}$ erg s$^{-1}$) but varying disc structures. From left to right: Original calculation by Owen et al. (2010); a reduction in dust temperature by $\sqrt{2}$; a reduction in the disc mass by a factor of 10. All figures shows very similar velocity and density structures in the wind.}\label{fig:diff_models}
\end{figure*}
This can be understood in terms of the model described in
    Section~2: the sound speed at the sonic surface scales as
    $c_s^2\sim\Psi_{\rm eff}\sim GM_*/2R[2-(z/R)^2]$, so that provided
    $z\ngg R$, order unity variations in height correspond to small
    variations in the effective potential (see
    Figure~\ref{fig:sonic_eff}). Thus the sound speed (and hence
    density and mass flux) at the sonic surface are insensitive to
    such changes in the height of the sonic surface. Since the
    mass-flux is predetermined then the sub-sonic flow structure
    adjusts itself to feed the sonic surface at this required
    rate. This results in a flow topology that is fairly independent of the temperature or density in the disc, even though it is ultimately these quantities that determine the actual location of the sonic surface.    
As expected we find that the total mass-loss rates and profiles are fairly invariant to dramatic changes in disc structure: the  total mass-loss rates for
the modified discs  range between $0.9-1.7\times10^{-8}$\msunyr compared to the original $1.4\times10^{-8}$\msunyr (Owen et al. 2010) for the same
 X-ray luminosity.
\subsubsection{X-ray attenuation} 
{ The basic theoretical outline described in Section~2 assumed that
  the heating at the sonic surface is dominated by optically thin X-ray photons. In the numerical models we also adopt an optically thin X-ray heating model with a hard column cut-off at $N=10^{22}$ cm$^{-2}$. In order to assess the consequences of these simplifications, we have performed additional simulations where the ionization parameter is instead given by:
\begin{equation}
\xi=\frac{L_Xe^{-\tau}} {n r^2}
\end{equation}  
In the case of a monochromatic X-ray source, this would simply replace
the X-ray flux with its value attenuated by the appropriate optical depth
at that energy. However, in reality we have a broad range of
energies, and
it is less clear how to attenuate the flux - in practice the attenuation
factor will reflect that for the photons that dominate the heating
locally.
Use of the ionisation parameter-temperature relation sidesteps
the
need to follow the propagation of photons of different energy and we thus
do not have the information to model the attenuation properly. However, we can
explore the consequences of attenuation by toy models in which the
optical depth of unity is fixed to three characteristic
columns $N=10^{20}$, $10^{21}$, $10^{22}$ cm$^{-2}$. The models with
X-rays attenuated at columns densities of $10^{21}$ and $10^{22}$
cm$^{-2}$ show little difference in structure and mass-loss rate to
that performed by Owen et al. (2010). However, the model with
attenuation at a column density of $10^{20}$ cm$^{-2}$ is unable to
heat the outer regions of the disc and the flow is restricted to the
inner $R\apprle 10$ AU, resulting in a significantly reduced mass-loss
rate. We can understand this result by inspecting Figure 9, which superposes
contours of equal column density to the source on a temperature map
in the case of a model without attenuation. We see that the contour
with a column to the star of $10^{21}$ cm$^{-2}$ lies close to the sonic surface
and thus (given the arguments that we have given here about the
importance of the sonic surface in setting the mass flux) we expect that
attenuation at column densities of $10^{21}$ cm$^{-2}$ or above will not greatly
affect the mass flow.
}
\begin{figure}
\centering
\includegraphics[width=\columnwidth]{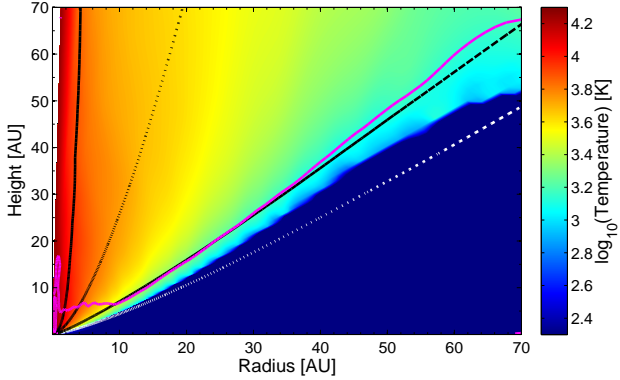}
\caption{Temperature structure of the wind from a star with an X-ray luminosity of $L_X=2\times10^{30}$ erg s$^{-1}$ and a stellar mass 0.7\msun. The solid magenta line shows the sonic surface, dot-dashed line a column of $10^{22}$ cm$^{-2}$, the dashed line a column of  $10^{21}$ cm$^{-2}$, the dotted line a column of $10^{20}$ cm$^{-2}$ and the solid black line a column of  $10^{19}$ cm$^{-2}$. Cold regions at columns $<10^{22}$ cm$^{-2}$ indicate regions where the dust temperature is higher than the temperature given by X-ray heating.}\label{fig:col_temp}
\end{figure}

{  We now need to ask whether the  extreme attenuation model (i.e. that
with an attenuation column of $10^{20}$ cm$^{-2}$, which does suppress
the mass-loss) is physically reasonable. Such an attenuation column
would be appropriate to photons of energy $\sim 0.1$ keV. Thus to argue
that $10^{20}$ cm$^{-2}$ is the relevant column density is to
argue that photons of higher energy (i.e. $> 0.1$ keV) are minor
contributors to  heating  gas to the temperatures
($3000-5000$K)  in  the region of the sonic
surface that dominates the mass-loss. 

The most convincing argument that the higher energy photons ($\apprge 0.3-0.4$ keV) are dominating the heating to temperatures $\apprle 5000$K is provided by the exercise conducted in Owen et al. (2010). In this exercise, the wind structure of a high density ($L_X=2\times10^{30}$ erg s$^{-1}$) unattenuated steady state
wind (obtained from a hydrodynamic calculation using an ionization parameter approach) was fed in to the {\sc mocassin} Monte Carlo radiative transfer code (Ercolano et al. 2003,2005,2008), which self-consistently
follows the attenuation
and propagation of X-ray photon packets as a function of energy. The results of this exercise, shown in Figure~9 of Owen et al. (2010), confirmed that the agreement between
the parametrised temperatures used in the hydrodynamics and the
temperatures generated by  {\sc mocassin} is excellent. We infer from this that it is higher energy photons ($\apprge 0.3-0.4$ keV) that are mainly responsible for heating the region of the sonic surface that dominates the mass-flux to the required temperatures ($\apprle 5000$K).
}
 
\subsection{Simulation Results: inner hole discs}
\begin{figure}
\centering
\includegraphics[width=\columnwidth]{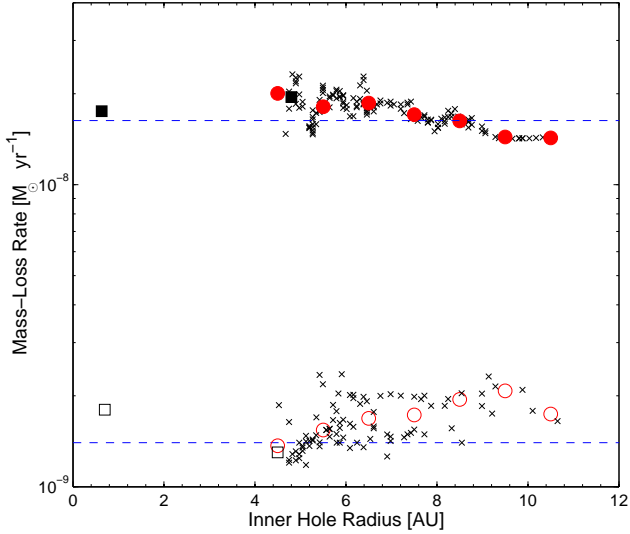}
\caption{Figure showing the mass-loss rate as a function of inner hole size for discs around a 0.1\msun star. The filled points are those for an X-ray luminosity of $2\times 10^{30}$ erg s$^{-1}$ and the open points are those for an X-ray luminosity of $2\times10^{29}$ erg s$^{-1}$. The crosses are the mass-loss rates calculated from the eroded inner hole models, while the circles are the moving average calculated from these eroding  mass-loss rates and the squares are the mass-loss rates from simulations which are maintained in steady-state. The dashed line shows the mass-loss rates for primordial discs with the same X-ray luminosity. }\label{fig:mdotih}
\end{figure}

In Figure~\ref{fig:mdotih} we show the mass-loss rates as a function
    of inner-hole size for the low mass stars (0.1\msun) compared with
    previous results of the higher mass-stars. 

As in the simulations reported in Owen et al. (2011b) we find that the mass-loss rate is approximately independent of the inner hole size. Figure~\ref{fig:mdotih} also shows that the steady-state simulations (squares) give identical results to the eroding simulations, giving confidence that the mass-loss rates obtained from the eroding simulations are accurate over the range of radii considered. The simulations show as expected that the mass-loss rate scaling is also linear with X-ray luminosity in agreement with the theoretical
arguments presented in Section~2 and with previous simulations.
\begin{figure}
\centering
\includegraphics[width=\columnwidth]{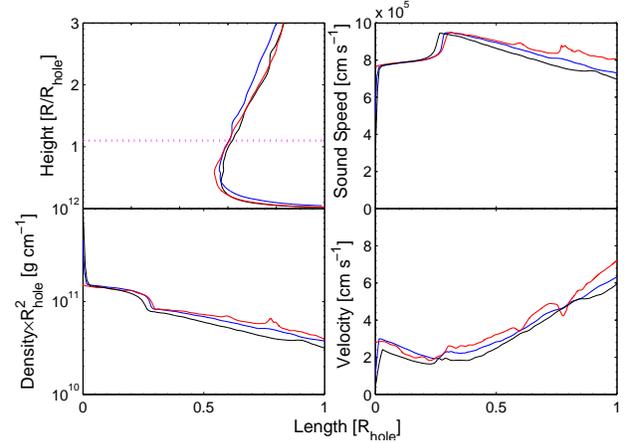}
\caption{Figure showing the streamline topology and density, sound speed and velocity along the sub-sonic section of the streamline. The streamline topology is shown in the top left hand panel scaled in terms of $R_\textrm{hole}$, with the sonic surface shown as the dotted line. Then bottom left panel show the density $\times$ $R_\textrm{hole}^2$, the top right show the sound speed and the bottom right show the velocity as a function of distance along the streamline 
normalised to $R_{hole}$. The black lines are for a hole of 14 AU, the blue line for a hole of 21AU and the blue for an inner hole of 30 AU.}\label{fig:hole_compare}
\end{figure}

\subsubsection{Understanding the Inner hole Scalings}
Examination of the streamline topology
for various inner hole radii (over a factor of $\sim 2$)  suggests that the
topology of the innermost streamline is approximately scale free (i.e
with an over all radial scale that varies with $R_{\textrm{hole}}$);
in  consequence the variation
of temperature and velocity along this streamline (as a function of
distance along the streamline scaled to $R_{\textrm{hole}}$) is approximately the same
in each simulation (see Figure \ref{fig:hole_compare}). 
{ This encourages us to construct a
scaling argument even though in the Appendix we show  models with different $R_{\textrm{hole}}$ cannot be
exactly self-similar. There is no {\it a priori} reason one would expect the temperature structure of these flows to be identical: this only follows if we 
use the numerical result that the streamline topology is fixed.
We argue in the Appendix that the fact that the relationship between the ionization parameter and temperature is not a simple power law implies that the conditions along two topologically identical streamlines can only be scaled to each other if the variation of temperature with (scaled) distance along the streamline is the same in both cases. Then the mass-flux scales as $n(l)A(l)$. For fixed temperature Equation~\ref{eqn:den_s} implies $n\propto R_{\rm hole}^{-2}$ and the stream-bundle area scales as $A\propto R_{\rm hole}^2$. Therefore, along this streamline the mass-flux is fixed and does not scale with inner hole radius.} 

The origin of a scalable innermost streamline is likely to originate from the fact that unlike primordial discs (where the thermal pressure is dominant in determining the streamline),  in this case the centrifugal force is important
since  material that is stripped from the rim of an inner hole
initially moves inwards.  As the effective potential is scalable in terms of $R_\textrm{hole}$ then it follows that (in the limit of negligible pressure)
the streamline topology would also be scalable in terms of $R_\textrm{hole}$. Provided the total mass-flux is dominated from the inner hole region - as the simulations indicate - then the total mass-loss should scale in a similar fashion to the mass-flux on the innermost stream-bundle, resulting in a total mass-loss rate that is approximately independent of the inner hole radius.

\section{Numerical tests: including `FUV-type' heating}
In Section~3 we argued that at radii $< 100$ AU the flow morphology is likely to resemble that shown in Figure~\ref{fig:X-rayfuva} where the mass loss rate is set entirely by the conditions at the X-ray sonic surfaces and where any FUV heating deeper must produce a flow that self-adjusts to supply the correct mass flux at the X-ray sonic surface.  In order to test whether the flow topology resembles that shown in Figure~\ref{fig:X-rayfuva} we perform some simulations that roughly aim to include the effect of extra-heating below the X-ray dominated region. If FUV heating was important to the calculation of photoevaporation rates then one would require a detailed calculation of X-ray and FUV heating and cooling, along the lines of combining the X-ray algorithm with an FUV algorithm (e.g. Richling \& Yorke, 2000). While such a calculation would be necessary to predict line emission from the warm atomic and molecular layer, we are here interested
in the simpler question of whether a plausible
modification of the structure at the
base of the flow due to the effect of FUV heating
will influence the flow at the X-ray sonic
surface. Therefore, a full X-ray +UV calculation is unnecessary and beyond the scope of this work and we instead employ a simple
prescription for FUV heating at the base of the flow.

We calculate the FUV flux throughout the disc/wind system in an identical manner to Richling \& Yorke (2000), including opacity only from dust extinction (e.g. we ignore opacity arising from the ionization of Carbon and Sulphur,
thus  overestimating the flux and gas temperature). We then turn the attenuated FUV flux into a temperature using a parametrisation of the previous  FUV calculations of Richling \& Yorke (2000) and FUV heating calculation of Gorti \& Hollenbach (2008): 
\begin{equation}
T_{\textrm{FUV}}=1400\textrm{K}\left(\frac{f_{\textrm{FUV}}}{300 \textrm{ erg s}^{-1}\textrm{ cm}^{-2}}\right)^{0.35}\left(\frac{n}{5\times10^6{\textrm{cm}^{-3}}}\right)^{-0.25}
\end{equation}
where $f_\textrm{FUV}$ is the attenuated FUV flux. This
parametrisation reproduces the temperature structure of the region directly irradiated by FUV in Richling \&  Yorke (2000) (their model A) and the
temperature structure listed for the `launching point' of the
calculations of 
Gorti \& Hollenbach (2008).
We perform three simulation with a large FUV luminosity of
    $1\times10^{31}$erg s$^{-1}$ { where we vary the FUV
    extinction}, and choose a depleted (from ISM type dust) value of
    the FUV extinction of 300 cm$^{2}$ g$^{-1}$ used by Richling \&
    Yorke (2000) and a extinction that is depleted by a further factor
    of 10 to 30 cm$^{2}$ g$^{-1}$. Furthermore the calculations of
    Gorti et al. (2009) indicate that more depleted discs drive more
    efficient flows (as in their calculations the mass-loss remains
    constant as the accretion rate - and thus FUV luminosity -
    falls). Therefore, we perform a calculation using the low mass
    disc described earlier ($M_d=2.6\times10^{-3}$\msun) and use an
    FUV extinction of 2 cm$^{2}$ g$^{-1}$ indicative of a very heavily
    settled disc atmosphere (D'Alessio et al. 1998,1999).  At every
    point in the disc/wind system we choose the heating mechanism
    (X-rays, FUV or dust) that produces the highest temperature, where
    we note the dust temperatures used already include the effects of
    FUV heating from accretion luminosity as well as the chromospheric
    contribution (D'alessio et al. 2001). { Furthermore, as we
    attenuate the FUV flux there is no need to apply a column cut-off
    as in the X-ray case. Thus provided the FUV gas temperature is
    higher than the X-ray or dust temperature at any point in the grid
    (even those with $N_H>10^{22}$ cm$^{-2}$) it can be heated by the FUV radiation}.  The FUV extinction is fixed throughout the calculation domain: this is reasonably valid as Owen et al. (2011a) showed that photoevaporative winds are capable of entraining small dust particles, which will dominate the opacity at FUV wavelengths.  We emphasise that  this is no attempt to calculate the self-consistent thermal structure of a disc heated by FUV/X-rays, but purely an investigation into the influence of  FUV type heating at the base of an X-ray heated
flow.  We demonstrate this further by running calculations in which we vary the parametrisation; while this of course has no physical basis it is purely a simple way to demonstrate that are our results are not being driven by our detailed
assumptions about FUV heating.
We thus calculate an extra 4 models for the maximum FUV penetration model, where we vary the power law indices (in Equation~12 in all four possible combinations) of the flux and density variation by factors two in both directions.

{ Finally, we investigate whether the X-ray to FUV
    luminosity ratio has any bearing on the results (particularly any
    transition from the flow topology shown in Figure 3(a) to Figure 3(b) when the X-rays are weak compared to the FUV luminosity). We thus fix the FUV luminosity at the large value of $1\times10^{31}$ erg s$^{-1}$ and systematically lower the X-ray luminosity to the lowest values observed $\sim 10^{28}$ erg s$^{-1}$.}

\subsection{Results}

Our calculations using this extra `FUV-type' heating does not affect the mass-loss rates obtained ({for $L_X=2\times10^{30}$, $L_{\rm FUV}=1\times10^{31}$ erg s$^{-1}$)}: we find   a value of $1.2\times10^{-8}$\msunyr for the initial calculation with the same depleted FUV extinction coefficient as Richling \& Yorke (2000),  a value of $1.1\times10^{-8}$\msunyr for the calculations with the very depleted FUV extinction coefficient, and a value of $1.3\times 10^{-8}$\msunyr for the mass depleted disc. In all simulations we find as expected that the sonic surface occurs in the X-ray heated regions and we therefore confirm the simple ideas presented in Section~2,  as well as demonstrating that the photoevaporation rates are controlled by the X-ray physics alone.

\begin{figure}
\centering
\includegraphics[width=\columnwidth]{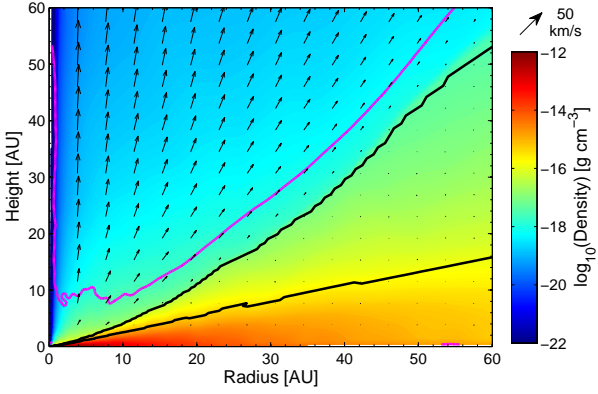}
\caption{Velocity and Density structures of winds including an extra `FUV' like heating mechanism, the arrows show the velocity in the wind and the black lines show regions where the extra heating mechanism is dominating over the X-rays and dust and the magenta contour shows the sonic surface. The model shown is the depleted disc with the extremely low FUV extinction allowing for maximum penetration of the underlying disc.}\label{fig:FUV_test}
\end{figure}

 In Figure~\ref{fig:FUV_test} we show the density and  velocity structure of
the maximum FUV penetration model (i.e. the mass depleted disc).
This shows as expected that the sonic surface occurs in the X-ray heating region, while the FUV heated flow is confined to a layer between the bound dust heated disc and the X-ray heated flow (i.e. between the two black lines in
Figure ~\ref{fig:FUV_test}). We further find that even though the density and temperature structure of the `FUV' heated region varies between model runs, in all cases the thermal structure of the sonic surface is as shown in Figure~\ref{fig:sonic_srf}. Therefore, the FUV layer must be adjusting its structure in order to `feed' the X-ray heated region with the correct mass-flux, and satisfy the hydrodynamical conditions at the sonic surface. Therefore, extra heating below the X-ray heated region has little effect on the photoevaporation rates in the range $R\sim 1-100AU$. 

\begin{figure}
\centering
\includegraphics[width=\columnwidth]{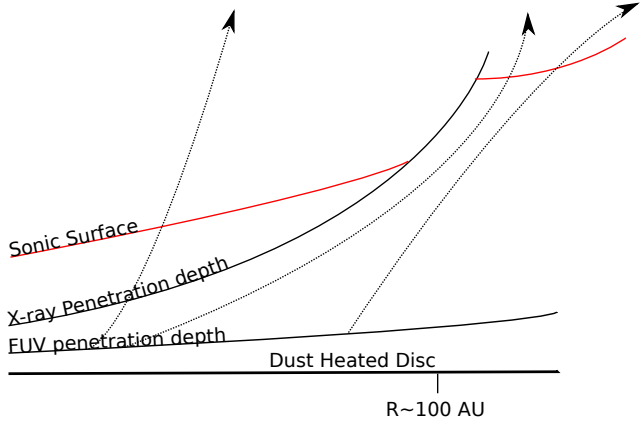}
\caption{Cartoon of an X-ray flow with an FUV wedge flow underneath.}\label{fig:prim_flowc}
\label{fig:X-rayfuvc}
\end{figure}

The structure of the system shown above does allow in principle for some
additional mass loss due to FUV heating if some of the gas between the two black
lines flows radially outwards through the FUV heated wedge rather
than feeding material to the X-ray sonic surface. If such a `wedge flow'
(see schematic depiction in Figure~\ref{fig:prim_flowc})'
did exist then (given the temperature of the FUV heated gas) it would
not undergo a sonic transition until a radius of order $100$ AU, i.e. close
to the outer boundary of the grid. We find that gas does leave the grid sub-sonically in the FUV heated region. However, due 
to the use of outflow boundary conditions on this grid boundary, it is impossible to tell if this flow is real or an artefact of outflow boundary conditions (which are only exact for super-sonic flow). Therefore, we are unable to further hypothesise on the possible existence of such  FUV `wedge' flows.
\subsubsection{Effect of X-ray luminosity}
{ In Figure~\ref{fig:FUVLx} we show how the mass-loss rates vary with decreasing X-ray luminosity, compared to the mass-loss rates expected from pure X-ray models. For $L_X > 10^{29}$ erg s$^{-1}$ the flow
resembles that of Figure~12: the X-ray sonic surface extends out
to cylindrical radii $\apprge$ 50 AU and sets the mass loss rate from the disc.
At lower X-ray luminosities (for a fixed $L_{\rm FUV}$ of $10^{31}$ erg s$^{-1}$) the
subsonic FUV region puffs up vertically and drives the X-ray sonic surface
and
upper region of the FUV heated region almost vertical at a radius
of $\sim 10$ AU. This means that the predominant mass loss in the disc
becomes at this point dominated by sub-sonic flow from the FUV region.
As noted above, we do not trust the mass loss rates that we obtain
in the case that material leaves the grid  sub-sonically so that the
mass loss rates at the low $L_X$ end of Figure 14 should not be taken
as a good measure of the total mass-loss rates. Thus considerably
more investigation is required to describe photoevaporation in the
limit of low X-ray to FUV luminosity ratios.

Instead, the chief result
of Figure 14 is that the flow morphology and mass-loss rates are X-ray
dominated provided that $L_X/L_{\rm FUV}\apprge 0.01$.  While stars with X-ray luminosities $<10^{29}$ erg s$^{-1}$ only represent a small fraction $\sim 5-10\%$ of the population of young stars (G\"uedel et al. 2007; Owen et al. 2011b), the role of combined FUV/X-ray heated flows will certainly be important for these objects if the FUV luminosity continues to remain large ($\sim100L_X$) throughout the disc's lifetime.

\begin{figure}
\centering
\includegraphics[width=\columnwidth]{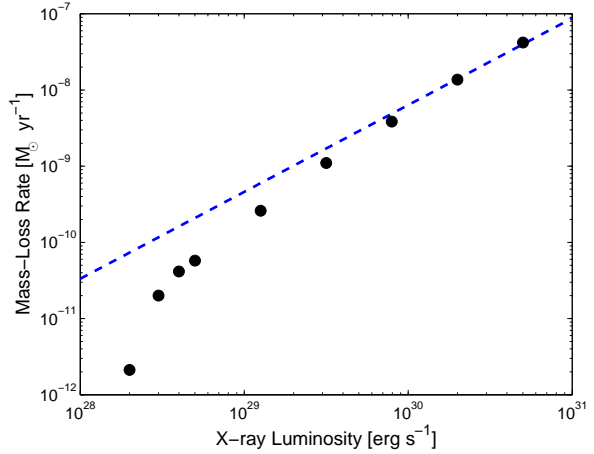}
\caption{Mass-loss rates arising in  the X-ray heated flow from models that include the FUV heating parametrisation, plotted as a function of X-ray luminosity. The dotted line indicated the mass-loss rate expected from a pure X-ray model. See text for a discussion of the reliablity of the results in the limit of low X-ray luminosities.}\label{fig:FUVLx}
\end{figure}
\section{Final Clearing of the Disc}\label{sec:sweep}
\begin{figure*}
\centering
\includegraphics[width=\textwidth]{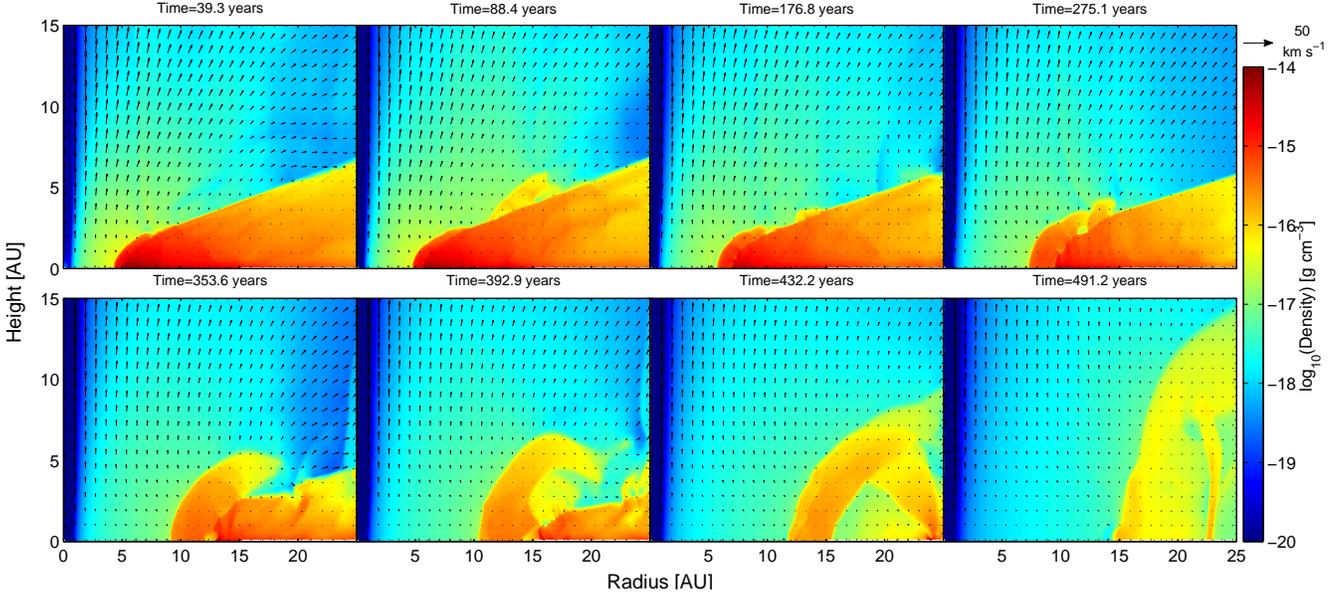}
\caption{Time evolution of a simulation of a disc with an eroding inner hole and an X-ray luminosity of $2\times10^{30}$erg s$^{-1}$ around a 0.1\msun star. The colour map shows the density structure and the arrows indicate the velocity structure. Each panel indicates a snapshot of the flow structure. }
\label{fig:hole_time}
\end{figure*}
In Figure~\ref{fig:hole_time} we show the evolution of a simulation with $M_*=0.1$\msun \& $L_X=2\times10^{30}$ erg s$^{-1}$ in which the disc hole is eroded during the course of the simulation (and for which the mass-loss rate during the evolution is shown by the crosses in Figure~10). During the first 300 years of evolution,
 the photoevaporative flow slowly strips off gas from the inner edge of the 
disc. During this stage, there is a {\it radially thin} layer of
bound  X-ray heated gas ($T_X\sim 200-400$K) on the inside rim of the disc and the flow
direction of the escaping gas is radially {\it inwards}. The origin of this layer is fairly easy to understand: as the fixed streamline topology suggested by the simulations fixes the density and structure in the flow, the flow cannot by construction consist of a fixed column density. It turns out that the column density of the flow is much less than $10^{22}$ cm$^{-2}$. Therefore, the X-ray heated region must contain not only the flow but also a warm bound X-ray heated region, that is separated from the flow by a contact discontinuity.   
 As the hole grows, the radial width of the layer of warm bound X-ray heated gas grows in size in order to still maintain the critical column
density for X-ray heating of $10^{22}$ cm$^{-2}$.  Once the hole size grows to $R_{\textrm{hole}}\sim 10$AU, the radial width of this region
becomes comparable with the vertical scale height of the X-ray heated gas.
At this point,  the topology of the  X-ray heated gas just within the rim changes
from nearly radial inflow to one where there is a strong (roughly sonic)
expansion velocity {\it normal to the disc plane}. This is seen in Figure
\ref{fig:hole_time} as the plume at the inner edge of the outer disc which,
in the lower panels, rolls over to envelope the  disc. The importance of this
change of flow topology is that it allows gas to be evacuated out of
the line of sight between the X-ray source and the disc rim. 
This thus  increases the penetration
of X-rays in the mid-plane. 
The panels at $353.6$ years and $392.$ years illustrate the formation of
a low density  bubble of heated gas between the expanding plume and the residual disc.
One can also discern that this bubble is expanding radially into the
disc and compressing the disc's inner rim. We term this effect
`thermal sweeping' and note that it operates on a roughly
dynamical time-scale (i.e.  $\sim$100 years at 10AU) which is not much more
than the free expansion time of X-ray heated gas. We emphasise that
this evolution is much faster than what occurs in the early stages
of clearing inner hole discs, when the erosion rate is limited by the fact
that the flow is radial and hence the evaporating gas continues to
provide a degree of shielding to the disc's inner rim. 

 This `thermal sweeping' was not seen in the simulations calculated in Owen et al. (2010,2011b) for discs around a $0.7 M_\odot$ star. We can understand this
difference  in that our simulations have a constant initial disc to star
mass ratio and the previous simulations therefore had higher disc masses
and higher mid-plane densities.
This implies that the critical column for X-ray absorption was attained over
a correspondingly smaller radial length scale;  thus, at a given radius,
the ratio of radial thickness to vertical scale height of the X-ray
heated gas is smaller in the
simulations with higher disc mass. These simulations therefore remained 
in the regime where the stripped gas moved radially inwards;  we
expect that (if we had the computational resources to calculate inner hole 
models around higher mass stars  with very large radii) we would
at some stage have entered the regime where the heated layer became 
radially thick
and that `thermal sweeping' would  ensue in this case also.

When this X-ray warm, bound region is small it must be in dynamical balance with the photoevaporative flow and the dust heated disc. The time-scale on which it can obtain this equilibrium is the sound crossing time of the region $t_\textrm{sc}\approx\Delta/c_s$, where $\Delta$ is the width of this region. When $\Delta/H\ll 1$ (where $H$ is the vertical scale height), this region can adjust to dynamical equilibrium quickly and  the disc/flow system is stable, even though the pressure in the region is significant. However, at the point when $\Delta\sim H$ this region cannot rapidly obtain dynamical equilibrium, and therefore will expand vertically on the same time-scale. This vertical expansion can clearly be seen in the time series plots, and as it expands the mid-plane density drops, resulting in higher temperatures. This in turn accelerates the vertical dispersal and a runaway ensues. Eventually, a high pressure region develops at the edge of the dust heated disc that pushes the remaining disc material away rapidly on the dynamical time-scale. 

In order to model this process we use the fact that the warm bound X-ray heated region must be in dynamical equilibrium with both the photoevaporative flow and the dust heated disc. The pressure is largest where bound X-ray heated region meets the dust heated disc, this region is highly sub-sonic and the requirement of dynamical equilibrium is simply one of pressure equilibrium. Thus one may write:
\begin{equation}
k_Bn_{\rm X}T_{\rm X}=P_{\rm dust}\label{eqn:pressure}
\end{equation} 
where $n_X$ is the density of the bound X-ray heated region. Now we can use Equation~\ref{eqn:pressure} to find $\Delta$:
\begin{equation}
\Delta=\frac{N_X}{n_X}
\end{equation}
where $N_X$ is the X-ray penetration depth (i.e. $N_X=10^{22}$ cm$^{-2}$). Thus the requirement for this process to begin $\Delta=H$, simply becomes:
\begin{equation}
P_{\rm dust}(R)\le\frac{N_X k_B T_X}{H(R)}\label{eqn:sweep}
\end{equation}
or :
\begin{equation}
\Sigma(R) \apprle \sqrt{2\pi}\,\mu m_h N_X \sqrt{\frac{T_X}{T_{\rm dust}}}
\end{equation}
{ Substituting standard values into this equation we find:}
\begin{equation}
\Sigma(R) \apprle 0.43\,\textrm{g cm}^{-2}
    \left(\frac{\mu}{2.35}\right)\left(\frac{T_X}{400{\rm
    K}}\right)^{1/2}\left(\frac{T_{\rm dust}}{20 {\rm
    K}}\right)^{-1/2}
\end{equation}
Thus this process will begin when the mid-plane pressure of the dust heated disc has dropped to a sufficiently small value, mainly through a reduction in the mid-plane density. This can be achieved either on account of a low disc mass or because the hole has grown to sufficiently large radius. This is shown in Figure~\ref{fig:sweep}
\begin{figure}
\centering
\includegraphics[width=\columnwidth]{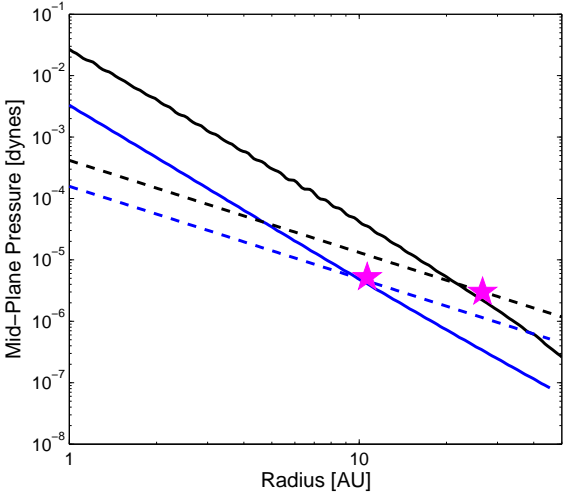}
\caption{Plot of the LHS (solid- pressure of dust heated disc) and RHS (dashed-critical pressure of X-ray heated region) of Equation~\ref{eqn:sweep}, to estimate the inner hole radii at which thermal sweeping begins. The stars show the radii the simulations indicate this process begins. Black represents a disc around a 0.7\msun~ star while blue shows a disc around a 0.1\msun (shown in Figure~\ref{fig:hole_time}).}    \label{fig:sweep}
\end{figure}  
where in order to test the applicability of Equation~15 we also perform a
    re-simulation of model F from Owen et al. (2010) (a disc with an
    initial hole of ~20 AU, an X-ray luminosity of $2\times10^{30}$
    erg s$^{-1}$ and mass of 0.7\msun), but the mid-plane pressure has
    been reduced to the point where Equation~\ref{eqn:sweep} predicts
    this process will occur when the disc hole size modestly exceeds its initial value. This re-simulation shows similar
    morphology to that shown in Figure~\ref{fig:hole_time} and the
    disc is completely cleared via the `thermal sweeping' mechanism
    described above as the hole reaches $\sim 25$ AU (shown in black
    on Figure~\ref{fig:sweep}). { The remaining disc mass just
    before this process sets in is around 5-10 Jupiter masses, and
    therefore a small fraction of the disc's initial mass is
    cleared in this process: roughly $\sim 80\%$ of the disc's initial
    material in accreted onto the star, $\sim 10\%$ lost in a
    photoevaporative wind and the final $\sim 10\%$ disperses through
    this process of `thermal sweeping'.}  

However, it is worth mentioning that these simulations
 were designed to study photoevaporative flows in the optically thin X-ray heated gas. On the other hand,
once `thermal sweeping' is initiated, the flow becomes sonic close
to the location where X-ray heating is initiated and in this
case  a simple column cut-off will not accurately reproduce the thermal structure of this region. Furthermore, as the gas is considerably cooler (hundreds to thousands rather than thousands to tens of thousands K as in the photoevaporation calculations) we cannot neglect the effect of molecular cooling, along with the added effect of FUV heating.  This latter depends on the dust properties at the mid-plane of the disc at large radii: although the X-ray heated region
is probably opaque to FUV radiation in the case of dust that is ISM-like,
the FUV may penetrate beyond the X-ray heated region in the case of
dust that is significantly depleted. 
The necessary calculations (including  X-ray and FUV radiation along with dust evolution)  are certainly beyond the scope of this work. We here 
restrict ourselves to stating that our simplified calculations suggest
that the disc enters a `thermal sweeping' phase (which clears the
residual outer disc on roughly a dynamical time-scale)
once the disc enters a region when the mid-plane pressure is
    sufficiently small. 

{ Finally, while the instability sets in on
    dynamical time-scales, simple energetic calculations suggest it
    may take $\sim10$ dynamical time-scales (at the inner edge) to unbind the outer
    regions of the disc
    at the lowest X-ray luminosities. This is somewhat longer than our simulations would predict in this case, since the simulations - by imposing radiative equilibrium - neglect the effect of adiabatic cooling. Hence, adiabatic cooling may be
    important during the final stages of `thermal sweeping', but not
    during the initial stages (onset of the instability) as the amount of material contained in
    the bound X-ray heated region is small compared to the total
    disc mass. However,
    10 dynamical time-scales (at the inner edge, $\sim10^{4}$ years) is still considerably faster than any
    other gas removal process operating at this stage. 
}

\section{Implications for Disc Evolution \& Dispersal}
The ultimate goal of the photoevaporation theory is to be able to predict  disc evolution and  lifetime based on properties of the underlying systems. At its most basic level we have shown throughout this work that it is only stellar X-ray luminosity that sets the photoevaporation rate and hence is the major factor in determining the disc lifetime. We have argued that the vigorous photoevaporation rates ($10^{-8}-10^{-11}$ \msunyr) are an inescapable consequence of X-ray irradiation (provided X-rays reach the disc's surface, an issue discussed in detail in Section~\ref{sec:uncer}), Furthermore, we showed that variations in disc structure and the inclusion of EUV and FUV radiation fields have little effect on the flow structure and derived mass-loss rates provided $L_X/L_{\rm FUV}>0.01$. 

The major consequence of this model is of course that discs around stars (of the same mass) with higher X-ray luminosities lose their discs first. An obvious implication of such a result is that disc-less stars will have on average a higher X-ray luminosity than disc-bearing stars at the same age. This has been 
observationally recognized for a while (e.g. Preibisch et al. 2005)
and Owen et al. (2011b) showed that the X-ray photoevaporation model
is quantitatively consistent with these observations. Ingleby et al. (2011) has
attempted one of the first studies to investigate the role of high
energy irradiation on disc dispersal; unfortunately the available
X-ray data for the study was incomplete in both X-ray luminosity and
mass, meaning it was impossible to disentangle the effects of stellar
mass and X-ray luminosity on disc dispersal and the result was
inconclusive.  Now we have fully developed the photoevaporation model
down to lower masses, we would have been hopeful that it would be
possible to make predictions of disc evolution as a function of
stellar mass. However, as described in Section~\ref{sec:life}, the
disc's lifetime also depends on how the viscous time varies with
stellar parameters and speculation about its variation with stellar
mass is well beyond the aims of this paper. It is worth noting though
that as discussed in Section 8.1, sensible assumptions lead to the conclusion that
disc lifetime is not a strongly dependant function of stellar mass and
is likely to decrease only weakly with increasing stellar mass, in
agreement with the observations (e.g. Ercolano et al. 2011b).  

    \subsection{Transition Discs}\label{sec:tran}
	It was the observations of the frequency of `transition'
    discs (Strom et al. 1989; Skrutskie et al. 1990; Kenyon \&
    Hartman, 1995) that lead to the development of the
    photoevaporation mechanism. Since the original observations, the
    number of detected discs that are consistent with the original
    characteristics of a transition disc
    (i.e. a large drop in opacity at short wavelengths, while
    returning to primordial levels at longer wavelengths) 
    has increased. However, there is no clear
    observational or theoretical consensus as to  what a `transition' disc
    actually is and it is becoming clear that the population of observed
    `transition' discs are not a homogeneous sample (e.g. Alexander \&
    Armitage, 2009; Owen et al. 2011b). The observations now span a range that
    includes accreting, non-accreting objects and those with small
    amounts of gas and dust inside the opacity `hole';  it is unlikely 
    that one mechanism is responsible for all objects
    (Cieza et al. 2010).  Some `transition' disc samples are now being split into `transition' and `pre-transition' populations (e.g. Espaillat et al. 2010; Furlan et al. 2011).

The three main mechanisms that can account
    for the formation and structure of this class of objects are:
    photoevaporation, planet formation and grain growth. Owen et
    al. (2011b) showed that a large fraction ($\apprge50\%$) of solar
    type `transition' discs were consistent with the X-ray
    photoevaporation scenario. In the X-ray photoevaporation scenario
    a `transition' disc goes through two stages of
    evolution. Once the gap opens, the dust in the residual inner disc
    is rapidly (i.e on a time-scale much less than the viscous evolution
    time of the gas) depleted as a result of inward migration due to gas
    drag. The loss of dust (and its associated opacity) from the inner
    disc would result in the
    object being classified as a transition disc even while it
    still possesses an inner gas disc with its associated accretion.
    This
    phase lasts approximately
    $\sim10\%$ of the observable disc's lifetime. Once the inner disc has entirely
    drained onto the central star, the inner edge of the outer disc is
    directly exposed to X-ray irradiation and the outer disc is then
     eroded
    to large radius. During this phase a centrifugal barrier prevents
    material accreting from the inner-edge onto the star and the disc
    would appear as a non-accreting `transition' disc with a large
    inner hole. { The structural properties of observed `transition'/`pre-transition' disc are currently more detailed than the photoevaporation model can predict. This must await the development of coupled dust and gas evolution models that are correct when the dust-to-gas ratio is no longer small. Therefore, at this stage we do not distinguish between `transition' and `pre-transition' discs. } {Although discussed in Section~7.2 the inclusion of `thermal sweeping' significantly changes the accreting to non-accreting fraction predicted by previous photoevaporation models, to a population that will be a majority of accreting transition discs.} 

 The observational data however also contains examples of transition discs
with large inner holes ($R_{\textrm{hole}}\apprge 20$AU) and large
accretion rates ($\dot{M}\sim10^{-8}$\msunyr) that are simply
impossible to create through any kind of photoevaporation model. It
has been suggested that these objects could have their large accretion
rates set by FUV ionization and MRI activation at the inner edge of
the hole (Perez-Becker \& Chiang, 2011); 
in this case the inner hole (or gap)  is created by the
tidal effect of a planet within the hole and dust filtration at the inner
edge of the outer disc is invoked to reduce the opacity of the inner disc.
However, such models are at an early stage of development and is not
clear what kind of planetary system would be required to match the
accretion rates 
and spectral energy distributions in these systems.
\begin{figure}
\centering
\includegraphics[width=\columnwidth]{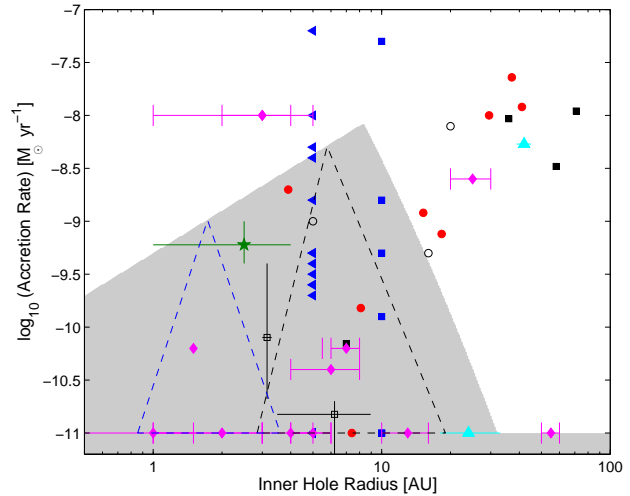}
\caption{The inner hole radius and accretion rate of low-mass objects
    classified as `transition' discs by varies authors: cyan
    triangles (Brown et al. 2009); blue filled triangles \& squares
    (Cieza et al. 2010); black open squares (Ercolano et al. 2009); black filled squares (Espaillat et
    al. 2008,2010); black open circles (Hughes et al. 2009, 2010); red
    filled circles (Kim et al. 2009); magenta filled diamonds (Merin
    et al. 2010); green filled star (Najita et al. 2010). Where
    error-bars or ranges are listed they are shown. Any object
    classified as non-accreting it is placed on the plot with an
    accretion rate of $10^{-11}$\msunyr.
The grey region represents the region of parameter space where the
    objects are consistent with gaps created by photoevaporation. See
    text for more details. The blue and black dashed regions represent those regions for a 0.3 and 1\msun~ star respectfully.}\label{fig:trans}
\end{figure}

Using the photoevaporation theory developed in the previous sections
    we can delineate  the region of transition disc parameter
    space within which we expect objects created by photoevaporation to 
    lie for stars with $M_*< 1.5$\msun (This exercise is similar to that undertaken by Owen et al. 2011b except that we have now extended the predictions to different stellar mass ranges and added new observational data points). Since the radial scale of the photoevaporative flow scales
    linearly with mass, the radius at which the gap opens will
    obviously scale linearly with mass too. Furthermore, since the gap
    opens when the accretion rate falls below the photoevaporation
    rate, then stellar mass effects are not important in the determining
    the accretion rates of transition discs other than through
    the implicit stellar mass dependence of the X-ray luminosity (and hence
    photoevaporation rate).  
    
     In
    Figure~\ref{fig:trans} we show the region in which
    photoevaporating transition discs are expected,  over-plotted with
    observations\footnote{Several other types of astrophysical objects
    are found to enter these samples (e.g. AGB star or debris discs): 
    in these cases we follow the definitions of the authors and only
    plot objects that are classified as protoplanetary discs with a
    gap/hole in its dust disc.}of discs classified as transition discs: cyan
    triangles (Brown et al. 2009); blue filled triangles \& squares
    (Cieza et al. 2010); black open squares (Ercolano et al. 2009); black filled squares (Espaillat et
    al. 2008,2010); black open circles (Hughes et al. 2009, 2010); red
    filled circles (kim et al. 2009); magenta filled diamonds (Merin
    et al. 2010); green filled star (Najita et al. 2010). We note that
    as Cieza et al. (2010) do not fit for an inner hole radius but
    rather list the {\it Spitzer} band of the inner hole. Therefore, we split the sample
    into two: the objects with a $\lambda_{\textrm{turnoff}}\le5.8\,\mu$m, the inner
    hole is conservatively set to be less than 5AU, for
    $\lambda_{\textrm{turnoff}}=8\,\mu$m we again conservatively
    set an inner hole of 10AU (since the inner-hole must occur in the
    temperature range $\lambda=8$ -- 24$\,\mu$m), we add that these are safe overestimates
    based on simple temperature structures of protoplanetary discs. It
    is also worth noting that at very low masses ($\sim0.1$\msun) 
    Ercolano et al. (2009) showed that current observations cannot 
    determine the presence of holes smaller that $\sim
    1$ AU, due to   a lack of contrast between the disc and the stellar photosphere.  

  The
    photoevaporation region is constructed by using the fact the photoevaporation profile is
    self-similar when the radius is scaled in terms of
    $R_g$. (i.e. if the gap opens at 1AU around a 1\msun star it would
    open at 0.1AU around a 0.1\msun star) and we use a X-ray
    luminosity scaling of $L_X\propto M_*^{3/2}$ (Preibisch et al. 2005,
    Guedel et al. 2007) to extend the region calculated by Owen et
    al. (2011b) to the full mass range. Again this shows that the observations can be separated into two groups, those with small inner-holes $\apprle 20$AU and small accretion rates $\apprle 10^{-8}$ \msunyr that are consistent with a photoevaporative origin, and those discs with large accretion rates and large holes that cannot have a photoevaporative origin.  
       
\subsubsection{Role of stellar mass}
Kim et al. (2009) noted that `transition' discs may show correlations between  disc properties and stellar mass. Within the X-ray photoevaporation framework, such correlation will be primarily driven by the variation of the X-ray luminosity with stellar mass ($L_X\propto M_*^{3/2}$; Preibisch et al. 2005,
    Guedel et al. 2007). In order to assess the role stellar mass may play in driving correlations between `transition' disc properties, we have extended the population synthesis study performed in Owen et al. (2011b) to lower (0.1\msun) and higher (1\msun) masses, where we adopt an initial disc mass that scales linearly with stellar mass. Furthermore, as the scaling of the initial viscous time with stellar mass is unconstrained (see Section 8.1), we preform runs where $t_\nu\propto M_*$ and $t_\nu\propto M_*^{-1}$, although the results shown are only mildly sensitive to the chosen scaling. 

\begin{figure}
\centering
\includegraphics[width=\columnwidth]{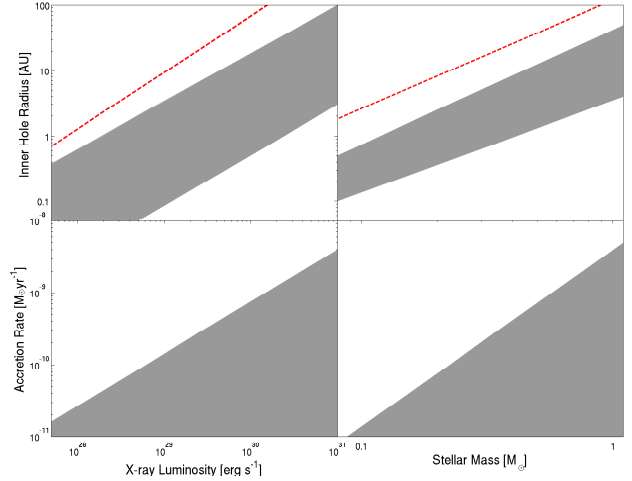}
\caption{Panels showing the predicted regions occupied by transition discs created by photoevaporation. The shaded region represents the parameter space occupied by accreting transition discs (i.e. those with $\dot{M}_*>1\times10^{-11}$\msunyr) , the dashed line represents the maximum radius a transition disc may reach (even when a non-accreting transition disc) before `thermal sweeping' sets in. }\label{fig:trancor}
\end{figure}    
     
     In Figure~\ref{fig:trancor} we show the predicted regions that the inner hole radius and accretion rates of transition discs occupy as a function of X-ray luminosity and stellar mass, both for accreting transition discs ($\dot{M}_*>1\times10^{-11}$\msunyr), and non-accreting transition discs before `thermal sweeping' sets in. In all cases they show strong positive correlations, driven by the radial scaling of the photoevaporating region with mass ($R_g\propto M_*$), and the positive correlation between mass and X-ray luminosity.  The shaded regions are calculated by randomly sampling in time the disc during the transition phase, and represent the area populated by all transition discs through the entire population's evolution. Thus, the correlations may vary with time for individual clusters, as only a certain fraction of discs will be in the transition disc phase at that point in time.

    \subsection{Relic Discs}
\begin{figure}
\centering
\includegraphics[width=\columnwidth]{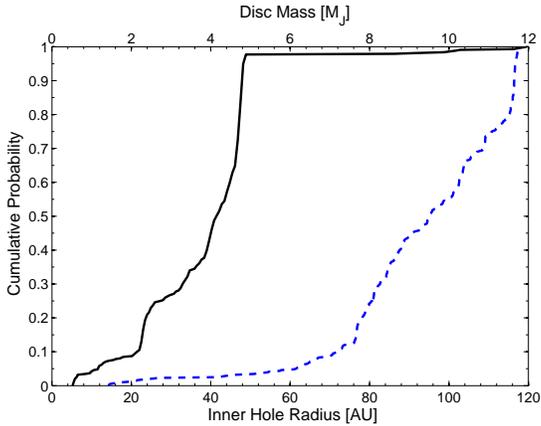}
\caption{Cumulative distribution of inner hole radii (solid line) and disc mass (dashed line) at which the thermal sweeping process starts for the disc models calculated in the Owen et al. (2011b) population synthesis model (0.7\msun stars).}\label{fig:relic}
\end{figure}
	As discussed in Section~\ref{sec:sweep} we have uncovered a further phase in disc dispersal, i.e. the rapid removal of disc material due to what we term
`thermal sweeping'. This phase begins at the point that the bound X-ray
heated gas just inward of the inner disc rim becomes geometrically thick
and when this gas starts to flow predominantly vertically, thus
allowing further X-ray penetration in the disc mid-plane.
The population synthesis calculation carried out in Owen et al. (2011b) of solar type stars ($M_*=0.7$\msun) did not include this effect and therefore predicted the existence of long lived relic discs, for a small fraction of the total population. These were discs with very large inner holes $R>100$AU and low photoevaporation rates; for objects with the lowest X-ray luminosities 
these relics were estimated to remain for  $~10$Myr and Owen et al. (2011b) argued that
such discs might well correspond to some sources that had previously
been classified as young (gas free) debris discs. This new thermal
sweeping effect now provides a mechanism for the removal of  these relic discs 
on a  rapid (dynamical)  time-scale ($\sim10^{3-4}$ years).   

  In order to investigate at what point we expect the thermal sweeping to
set in for the case of discs around higher mass stars (the calculations
reported in Section ~\ref{sec:sweep} were for stars of mass $0.1 M_\odot$) we apply the 
pressure criterion derived in Equation~\ref{eqn:sweep}, since this corresponds to the
point at which the radial width of the X-ray heated bound gas is $\sim H$.
We follow the population synthesis approach of Owen et al. (2011b), i.e. we
assume that all discs evolve according to a viscous similarity solution with
fixed parameters and where the only quantity that varies from star to star
is the X-ray luminosity  which sets the photoevaporation rate.
Then each star-disc system is evolved under the combined effects of
viscous evolution and photoevaporation, following  each system through
the phase of gap opening and outer disc clearing. We identify the point that `thermal sweeping' sets in with point
that the maximum mid-plane pressure in the viscous disc falls below the
value given by Equation~15. The variation of this radius with X-ray
luminosity is rather modest, given that X-ray luminosity affects the
radial profile of the clearing disc only through the mild effect
of the degree of depletion of the disc during the phase of photoevaporation
starved accretion (Drake et al. 2009; Owen et al. 2011b). Consequently  our population synthesis model
predicts that in the higher mass stars, thermal sweeping should set in
at a narrow range of radii around 40 AU and disc masses $\apprle 10 M_{\rm J}$ (see Figure~19).
\begin{figure}
\centering
\includegraphics[width=\columnwidth]{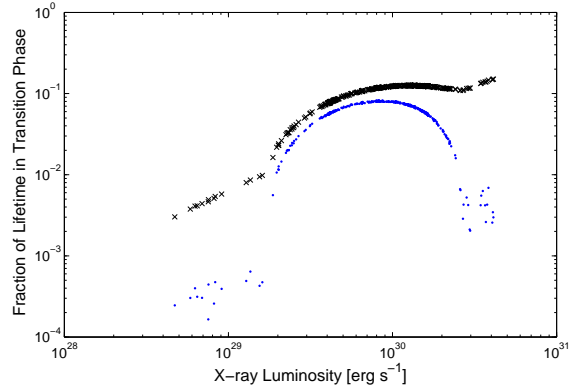}
\caption{Fraction of the disc's total lifetime that the disc presents as a transition disc (crosses) and a non-accreting transition disc (dots) as a function of X-ray luminosity.}\label{fig:tran_frac}
\end{figure}

{ Furthermore, the identification of this gas expulsion method
    significantly alters the ratios of accreting/non-accreting
    `transition' discs expected, since discs spend less time in the
    non-accreting phase than previously expected. In
    Figure~\ref{fig:tran_frac} we plot (as a function of X-ray
    luminosity) the fraction of the disc's lifetime that the disc
    spends as a transition disc (crosses) and as a non-accreting
    transition disc (dots). This indicates that discs with high
    ($\apprge10^{30}$ erg s$^{-1}$)  and low ($\apprle 10^{29}$ erg
    s$^{-1}$) X-ray luminosities spend the majority of their
    transition phase as accreting objects. Discs with intermediate
    X-ray luminosities spend similar amounts of time in the accreting
    and non-accreting phases. Thus over the entire population of
    discs, one would expect most transition discs to appear as
    accreting. We can understand the relative lack of
    non-accreting transition discs at both high and low X-ray luminosity as follows.
    At low X-ray luminosity the disc accretes most of its mass onto
    the star over its longer lifetime, so that the mid-plane pressure is
    low once it enters the accreting phase and thermal sweeping sets in rather promptly. At high X-ray luminosity
    the process of photoevaporation starved accretion (Drake et
    al. 2009; Owen et al. 2011b) is vigorous and once the disc enters
    the clearing phase the outer disc is very depleted of gas
    (compared to the result of pure viscous evolution). Again the mid-plane
    pressure is low and the disc enters the
    thermal sweeping phase quickly.  }

 The reason that such thermal
sweeping was not seen in our previous hydrodynamic simulations of
these more massive stars (Owen et al 2010) is that our hydrodynamical
calculations of discs with inner holes used input disc profiles that
were simply those of primordial discs truncated at a given value and did
not take into account the fact that the outer disc profile would have
been modified by viscous evolution and photoevaporation. The discs modelled
thus had higher pressures than discs that had been self-consistently
evolved and were thus less prone to thermal sweeping.  This is in
contrast to the sequence of disc clearing for a 0.1 \msun~ star
(see Figure 12) in which the evolution of the disc profile
is followed self-consistently in the hydrodynamic calculation. We
emphasise that our estimate that the disc is rapidly cleared in higher
mass (0.7 \msun) stars once the hole grows to $40$ AU is very
approximate given that it is not based on self-consistent hydrodynamical
simulation of disc clearing in this case.

\section{Discussion}\label{sec:discuss}
In this work we have indicated that one of the inescapable
    consequences of X-ray heating of a protoplanetary disc is a
    vigorous photoevaporative flow. The mass-loss rate is effectively
    determined by the X-ray luminosity  only (when $L_X/L_{\rm FUV}>0.01$), with disc structure and
    stellar mass only having weak effect on the resultant flow
    structure. This is easily understood in terms of the model
    presented in Section~2, where the conditions at
    the sonic surface are determined by the X-ray physics alone, so
    the disc below has to simply adjust to feed the sonic surface with the correct
    mass-flux. The consequence of such large mass-loss rates are that photoevaporation will be competing with accretion over much of the disc bearing life of young stars. Stars with higher X-ray luminosities will begin losing
    their discs first, followed by the lower X-ray luminosities stars
    at late times. Such an evolution is consistent with the
    observation that disc-less stars are systematically more X-ray
    luminous than stars still surrounded by discs (Neuhauser et al 1995; Stelzer \& Neuhauser 2001; Flaccomio et al. 2003; Preibisch et al. 2005). 

    We have shown it is extremely unlikely that the EUV 
    irradiation field plays any role in setting the photoevaporation
    rates: in any scenario where the EUV can reach the disc then so
    can the X-rays and the resulting X-ray driven wind is optically thick
    to the EUV. 
    We have also set out arguments (backed up by simple toy FUV heating models) 
that FUV heating becomes important only below the X-ray sonic surface\footnote{This is not to say that FUV heating is then irrelevant to setting the structure of the warm disc at intermediate heights (Gorti \& Hollenbach 2008,2009), but merely that is not the determinant of the mass-loss rate.}. Only when the X-ray luminosity drops to sufficiently low ($L_X/L_{\rm FUV}<$0.01) values does the FUV begin to shut off the X-ray flow, and may possibly drive a flow itself. This result raises the intriguing possibility
    that what determines the evolution of a disc (and presumably its
    ability to form planets)  is the star's intrinsic X-ray luminosity,
    rather than any variations in initial conditions when the
    star/disc system was created (Armitage 2011).  


\subsection{Disc lifetimes as a function of stellar mass}\label{sec:life}

  Since we have shown that the photoevaporation rate is independent
of stellar mass (at fixed $L_X$) then, given some knowledge of the 
statistical dependence of X-ray luminosity on stellar mass and of
the scaling of disc viscous parameters with  stellar mass, we can 
examine the predicted relationship between  disc lifetime and stellar
mass.  To this end we adopt the `null' disc    evolution method (Ercolano \& Clarke 2010; Owen et al. 2011b), i.e. we assume that
    discs evolve in a pure viscous manner until the accretion rate
    equals the photoevaporative wind rate. At this point the disc is
    cleared rapidly and we define this point as the disc lifetime. For
    a disc in which $\nu\propto R$ the viscous similarity solutions
    (Lynden-Bell \& Pringle, 1974) tell us:
    \begin{equation}\label{eqn:mdottime}
    \dot{M}_*(t)=\dot{M}_*(0)\left(1+\frac{t}{t_\nu}\right)^{-3/2}
    \end{equation}

   where $t_\nu$ is the disc's {\it initial} viscous time-scale (which,
for the assumed linear scaling of viscosity with radius) is itself
  just proportional to the disc initial radius ($R_1$).
    Thus we can simply write the disc life-time ($\tau_d$), when $t>t_\nu$ as:
    \begin{equation}
      \tau_d\propto\dot{M}_*(0)^{2/3}\dot{M}_w^{-2/3}t_\nu
      \end{equation}
    Now using the fact $M_d(0)/\dot{M}_*(0)\propto t_\nu$ and
    $\dot{M}_w\propto L_X$ we find
    \begin{equation}
      \tau_d\propto M_d(0)^{2/3}L_X^{-2/3}t_\nu^{1/3} 
\end{equation}

Observations and theoretical work that suggest the
    discs enter this viscous phase after a self-gravitating phase
    suggest that $M_d(0)\propto M_*$, Furthermore X-ray observations
    (Guedel et al. 2004, Preibisch et al. 2007, Albacete Colombo
    et al. 2007)
    indicate that $L_X\propto M_*^{3/2-5/3}$, adopting $L_X\propto
    M_*^{3/2}$ we find that:
    \begin{equation}
      \tau_d\propto M_*^{-1/3}t_\nu^{1/3}\label{eqn:disc_life}
    \end{equation}

  Unfortunately the scaling of $t_\nu$ (or equivalently $R_1$) with
stellar mass is unknown. Some suggested scalings have been motivated by
a wish to reproduce the claimed quadratic dependence of accretion rate
on stellar mass (although it is likely that this empirical correlation
- Natta et al. 2006 - may be largely driven by a mixture of sensitivity and
selection effects: see Clarke \& Pringle 2006, Tilling et al. 2008). One prescription
that yields this quadratic dependence is that of Alexander \& Armitage
(2006), 
who assume that most observed discs are at an age that is much less than
their initial viscous time-scale. Under this circumstance, the accretion
rate is simply linked to the ratio of initial disc mass to initial
viscous time-scale; a quadratic scaling of accretion rate on stellar mass
then demands that (assuming $M_d(0)\propto M_*$) the initial viscous 
time-scale
$t_\nu \propto 1/M_*$. If this were true, then Equation~\ref{eqn:disc_life} would
imply that disc lifetime scaled as $M_*^{-2/3}$.  
There are however a range of other possibilities based on reproducing
the claimed accretion rate versus stellar mass relation: e.g.   
if $M_d(0)\propto M_*^2$ and $t_\nu$ is independent of stellar mass
(as proposed by Dullemond et al. 2006) then $\tau_d \propto M_*^{1/3}$,
while if we adopt $M_d(0)\propto M_*$ and now assume instead that
discs are observed at times $ >> t_\nu$ then we would require
$t_\nu\propto M_*^2$ 
in which case we again obtain $\tau_d \propto M_*^{1/3}$.

 Clearly our estimates are resulting in a range of dependences
none of which has a strong stellar mass dependence. Given the
form of Equation~\ref{eqn:disc_life}, we see that a  strong dependence of dispersal
time-scale on stellar mass could only result if the initial disc 
viscous time-scale  were an {\it extremely}  strong function of stellar
mass. This would appear to be unlikely, both because there are no
obvious theoretical reasons to invoke such a strong mass dependence
and also because such a strong dependence would imply a large
stellar mass dependence of the relative numbers of stars observed
with discs
that are a substantial fraction of the mass of the central star. 

\subsection{Model Uncertainties}\label{sec:uncer}
It is worth spending some time dwelling on some of the uncertainties
    and possible sources of error in a direct application of the model
    to real objects. These can be divided into two categories: (i)
    assumptions that directly affect the calculated mass-loss rates;
    (ii) how the mass-loss rates affect disc evolution. 

\subsubsection{Uncertainties in mass-loss rates}
{ The analysis presented in Section~2 suggests that the mass-flux
    through the sonic surface is dominated by gas with temperatures
    around $2000-5000$K. At these temperatures, molecular cooling
    (which we neglected in the X-ray thermal calculation used here,
    which just consider line cooling) could in principle be important
    (Glassgold et al. 2004; Meijerink et al. 2008; Gorti \& Hollenbach
    2008,2009). In order to assess whether this is in fact the case
    (given that the densities at the sonic point are many orders of
    magnitude lower that those encountered in the preceding disc
    calculations) we need to estimate the relative contributions of
    the atomic line cooling at the sonic surface with that provided
    by molecules. From our calculations, the atomic line
    cooling.  is typically $\sim 10^{-13}$ erg cm$^{-3}$
    s$^{-1}$ at 2000K. Given the density at the sonic surface ($\sim 10^{6}$
    cm$^{-3}$),  then molecular cooling would become important only if
    the CO abundance ($X_{\rm CO}$) exceeded  of 0.1 at 2000K (using Equation~C7
    of Glassgold et al. 2004) which is many orders of magnitude higher than expected in discs (e.g. Bruderer et al. 2009). Thus we are confident that molecular cooling can be neglected at the sonic surface in photoevaporation calculations and hence the mass-loss rates will be unaffected. } 

While in this work and Owen et al. (2010) we went to great effort to
    indicate that the majority of factors that vary with time (e.g. disc
    structure, FUV/EUV luminosity) make very little difference to the 
    derived mass-loss rate, we made an important  assumption, namely the X-ray flux that is incident on the disc is an unvarying
    function of time. At first thoughts this is not a bad starting
    point since the time-averaged (X-ray variability due to flares are
    too fast to change a photoevaporative flow - Owen et al. 2011b)
    X-ray output of the star has
    been shown observationally  to be fairly constant as a function of
    time (Guedel et al. 2004, Ingleby et al. 2011), crucially what is
    important to the photoevaporation model is not the X-ray output of
    the star but the `soft' ($h\nu \apprle 1$keV) X-ray luminosity
    visible to the disc. 

    Firstly, since most X-ray studies are concerned with the total
    X-ray luminosity of the objects (normally taken out to $h\nu
    \approx 10$keV), not much attention has been paid to the
    evolution of just the `soft' component of the X-ray spectrum with
    time. While varying the amount of `hard' X-ray irradiation will
    not affect the heating and hence photoevaporation rate, it will
    have an effect on the translation between observed X-ray
    luminosity of an object and the total `soft' X-ray flux that
    reaches the disc. 

     A second factor is that various
    sources of absorbing column lie between the stellar corona and
    disc surface, for example: accretion columns, jets and (magnetic)
    disc winds may be present. Obviously if the absorbing column from
    these process approaches $10^{22}$cm$^{-2}$ then the X-rays will
    be screened from the disc and the mass-loss rates derived
    in the previous sections will be an overestimation of the actual
    photoevaporation rate.  Thus, simply assuming that the
    disc sees an X-ray source with the same luminosity as the
    intrinsic X-ray luminosity may be an oversimplification, leading to
    an over estimate of the photoevaporation rates. 

     Finally, although we are confident about the magnitude of the X-ray
     photoevaporation, we cannot rule out the possibility that beyond
     $100$ AU it is FUV photoevaporation that dominates the mass loss
     rate. This is because this is the distance at which the X-rays
     and FUV fields heat the gas to comparable temperatures and where
    also such temperatures are comparable with the local escape temperature.
    Gorti \& Hollenbach (2009) have suggested that this may be an important
    effect in truncating the disc at the outside (cf. a similar role for
    {\it external} FUV radiation in the case of the Orion proplyds Johnstone et al. 1998; Richling \& Yorke, 2000; Adams et
     al 2004). Such truncation would influence the viscous evolution of
     the disc (Clarke 2007) by imparting a different boundary condition
     from the one adopted in the viscous calculations as noted by Gorti et al. 2009.

\subsubsection{Uncertainties in disc evolution}

  Uncertainties in disc evolution {\it do not} affect the photoevaporation
rates for a given stellar mass, X-ray luminosity and (where relevant)
inner hole radius, given the insensitivity of these rates to the
structure of the underlying disc. However it is the interplay between
viscous evolution and photoevaporation that sets the sequence of 
gap opening and inner disc clearing. Viscous evolution sets the time-scale
on which the gap opens and, by setting the mass remaining in the
outer disc, also sets the time-scale of outer disc clearing.  

    Currently, the
    magnitude and even origin of the viscous processes in discs
 is not fully determined, nor the radial range within which various
   viscous processes are active 
    (see Armitage 2011 for an up-to date
    review). Therefore, in order to determine the evolution of discs
    we adopt a similarity law put forward by (Lynden-Bell \& Pringle,
    1974) wherein  $\nu\propto R$ since this crude assumption appears
    to be broadly compatible with observations of disc mass distributions
    and accretion rates in T Tauri stars
    (e.g. Hartmann et al. 1998). 

  Although the broad sequence of evolutionary phases outlined above is
 to be expected for any viscosity prescription, we are aware that there
 are quantitative effects that do depend on the magnitude and the
  radial dependence of the viscosity. In particular, we note that
    (as described in Section~\ref{sec:tran}), the inner disc
    rapidly becomes optically thin following gap opening,
     as it becomes significantly dust
    depleted. Such a change will undoubtedly change the thermal and ionization structure of the disc. This in turn can be expected to affect the viscosity
    in the disc with consequences, for example, for the predictions
   we have made concerning the relative incidence of
    accreting vs non-accreting transition discs
    with a photoevaporative origin.

\section{Conclusions}

In this work we have outlined the basic properties of the  photoevaporation of
discs around low mass stars by the 
X-rays from  the central  star. In this regime, the photoevaporation rates are set by the properties in the X-ray flow and are fairly insensitive to variations in disc structure and extra heating sources e.g. FUV radiation. 
At the most basic level this indicates factors that determine discs' lifetimes and evolution may stem from the stars' intrinsic X-ray luminosities. 
Our main conclusions are summarised below:

\begin{enumerate}
\item The photoevaporative mass-loss rates are primarily set by the X-ray heating alone, as the sonic surface always occurs within the X-ray heated region. Extra UV heating provides little effect on the determined structure of the X-ray wind and its associated sonic surface geometry  and  mass-loss rate  if $L_X/L_{\rm FUV}>0.01$.
\item We have derived from first principles the scaling of photoevaporative mass loss from primordial discs with X-ray luminosity and stellar mass and have
also derived analytical estimates of the total mass loss rates. The mass-loss rates scale approximately linearly with X-ray luminosity and have no explicit dependence on stellar mass. These scalings and absolute values have been
confirmed through a large suite of numerical simulations.

\item We do not predict a strong dependence of disc lifetime on stellar mass
for stars less massive than $1 M_\odot$, in agreement with observations.
Higher mass stars have higher X-ray luminosities but also have more massive
discs to be dispersed. The resulting time-scale for disc dispersal would
be a strong function of stellar mass only in the unlikely case that the
viscous properties of discs were an extremely strong function of stellar
mass.

\item We have explained why, in the case of discs with inner holes,
 the photoevaporative mass-loss rates scale linearly with X-ray luminosity and 
is independent of inner hole size and have calculated numerical models to calibrate the mass-loss rates.

\item We find that once the mid-plane pressure of a disc with an inner hole drops below a critical value (given by Equations~15 \& 16) the remaining  disc is cleared rapidly
(on a roughly dynamical  time-scale). This process (which we have termed
`thermal sweeping') sets in once the layer of bound X-ray gas on the
inside of the disc rim becomes thick (i.e. comparable with its vertical
scale height) since at this stage the heated gas escapes preferentially
normal to the disc plane and increases the exposure of the disc rim.
Such rapid dispersal removes the possibility of  the long lived
`relic' discs hypothesised in Owen et al. (2011b). We find that thermal
sweeping sets in at around $10-15$ A.U. in the case of low mass stars 
($0.1 M_\odot$) and we roughly estimate that the corresponding radius is
$\apprle 50$ AU for higher mass T Tauri stars ($\sim 0.7 M_\odot$).

\item We find that when we take into account the spread in masses and
X-ray luminosities of T Tauri stars, we can produce transition
discs that span a wide range of properties (hole radii and accretion
rate onto the star). In this way we can account for
a large fraction ($\apprge50\%$) of the observed transition discs. We
cannot, however, account for large holes with high accretion rates
(see Figure \ref{fig:trans}).

\item Photoevaporation and `thermal sweeping' ultimately destroy the final disc, but it is still accretion onto the central star that removes most of the original disc mass ($\sim 10 \%$ M$_*$): in most cases only $\sim 10-20\%$ of the original disc mass lost through in the processes of photoevaporation or thermal sweeping.

\end{enumerate}
\section*{ACKNOWLEDGEMENTS}
We are grateful to the anonymous referee, whose comments helped improve
    this work.
JEO acknowledges support of a STFC PhD studentship. This work was performed using the Darwin Supercomputer
of the University of Cambridge High Performance Computing Service
(http://www.hpc.cam.ac.uk/), provided by Dell Inc. using Strategic
Research Infrastructure Funding from the Higher Education Funding
Council for England. 

\begin{appendix}
\section{The theory of X-ray Photoevaporation: scaling relations}\label{sec:TheoryX}
 In this appendix we develop the theory of X-ray photoevaporation from discs
with a view to understanding how local quantities  vary as a function
of model input parameters.  We make use of the fact that in thermal equilibrium the 
temperature of optically thin,
X-ray heated gas is a roughly monotonic function of the ionisation parameter
$\xi=L_X/nr^2$ In qualitative terms, the flow starts highly sub-sonically at 
the
base of each streamline and is then accelerated by a mixture of pressure
gradients and effective gravity, $\bf g_\textrm{eff}$ (i.e. the combined effects of 
the gravity of the central star and a centrifugal term). It is assumed
that viscous effects are negligible over the flow; thus  the specific
angular momentum of each fluid element is conserved, being at
the Keplerian value at launch (cylindrical
radius $R_b$) with value $h^2=GM_*R_b$. Therefore, 

\begin{equation}
{\bf g_\textrm{eff}} = -{{GM_*}\over{r^2}} {\bf \hat r } + {{GM_*R_b}\over{R^3}}{\bf \hat R }
\end{equation}
where ${\bf\hat r}$, ${\bf\hat R}$  are unit vectors in the spherical radial 
and cylindrical radial  directions.

 In a steady state we can write the continuity equation
in the form:

\begin{equation}
 {{\dd {\log} \rho}\over{\dd l}} = -{{\dd {\log} u} \over{\dd l}} - {{\dd {\log} A} \over{\dd l}}
\end{equation}
(where $l$ is a coordinate measured along the streamline and  $A$ represents the area of a streamline bundle, so that
${{\dd {\log} A} /{\dd l}}$ is simply the divergence of the
unit vector along the streamline,  
${\bf \hat l} $). 
This allows us to write the steady 
state momentum equation along a streamline in the form:

\begin{equation}
\left[\left({{u}\over{c_s}}\right)^2 -1\right]{{\dd {\log} u}\over{\dd l}} = {{\dd {\log} A} \over{\dd l}} - {{\dd {\log} c_s^2}\over{\dd l}} + 
{{{\bf g_\textrm{eff}} \cdot {\bf \hat l} }\over{c_s^2}} \label{eqn:mom1c}
\end{equation} 
 Therefore a flow will undergo a sonic transition
at the point where the right hand side of Equation~\ref{eqn:mom1c} is equal to zero.
In the special case of an isothermal flow, we would recover the
result that in the absence of an external force ($g_\textrm{eff} = 0$) such
a transition occurs at the point of transition from converging to diverging
flow (the de Laval nozzle solution). Furthermore,  in the case of spherical outflow
under point mass gravity, the right hand side of Equation~\ref{eqn:mom1c}  yields the
sonic condition of a Parker wind (Parker, 1958).  
 
 However, the present case is much more complicated, since we do not know
the flow geometry {\it a priori}, because $g_{\rm{eff}}$ is non-negligible
at all radii and because $c_s$ is not constant. 
Although it is possible to solve for $u(l)$ (and hence $\rho(l)$ and $c_s(l)$)
for a fixed streamline topology, there is no guarantee that such a solution
would satisfy the steady state momentum equation perpendicular to the
    streamlines. We can use Equation~\ref{eqn:mom1c} to write the
    condition for the sonic surface as:
\begin{equation}
 c_{s}^2 = {{GM}\over{2R}}  \times {{R}\over{r}} \times {{f_g}\over{f_A}}
\end{equation}
where 
\begin{equation}
f_g = {{\bf-g_{eff} \cdot {\hat l}}\over{GM/r^2}}
\end{equation}
and
\begin{equation}
f_A = {{{\dd  {\log} A}\over{\dd l}} - {{\dd {\log} c_s^2}\over{\dd l}}\over{2/r}}
\end{equation}
The first two terms equate the temperature at the sonic surface to
the usual expression for a spherical Parker wind;  $f_g$ 
takes into account of the different component of the effective force
along the streamline compared with the Parker case (both due to the
centrifugal term and the non-spherical geometry), while $f_A$ takes
into account the different divergence of non-spherical flows and
also the fact the flow is non-isothermal. In thermal equilibrium the
    flow time-scale is long compared to any thermal time-scales and
    the derivative of the sound speed is small compared to the other
    terms (e.g. Adams et al. 2004). Furthermore, as the centrifugal
    force falls off rapidly with radius, $f_A$ and $f_g$ approach unity
    quickly (Begelman et al. 1983; Adams et al. 2004). Therefore
    provided the sonic surface is some distance from the base of the
    flow, the sound speed will be of order the Parker value.

At first look these remarks  appear to preclude further analytic
arguments,  but in fact we can make useful progress by deriving {\it scaling
relations}. First we consider the case where we have obtained a self-consistent
steady state flow structure (via a hydrodynamical simulation) for a given
set of input parameters, and  we then change some parameter of the
flow and obtain a new structure. We can then ask whether there are circumstances
where one case is simply a scaled version of the original case.

 If such a scaling does apply, then  every streamline in the first case
can be identified with a topologically identical streamline in the second case, which differs
only in its overall geometrical scale. Such a scaling also implies
that 
the variation of flow variables along each such streamline are scaled versions
of each other, 
i.e. we write
all variables in the form $u = u_0 \tilde u (\tilde l)$  where $\tilde l$
is the distance along the streamline scaled to the radius at the
base of the streamline ($R_b$) and $u_0$ allows for a simple re-scaling of the
velocity between the two streamlines. In this case we can write
Equation~\ref{eqn:mom1c} as: 
\begin{eqnarray}
\left[\left({{u_0}\over{c_{s0}}}\right)^2  
\left({{\tilde u}\over{\tilde c_s}}\right)^2 -1 \right] 
{{\dd {\log} \tilde u}\over{\dd \tilde l}} 
&=& {{\dd {\log} \tilde A} \over{\dd \tilde l}} - {{\dd {\log} \tilde c_s^2}\over{\dd \tilde l}}\nonumber\\ 
&&+ 
{{GM_*}\over{R_{b}c_{s0}^2 }} {{{\bf\tilde g_\textrm{eff}} \cdot {\bf\hat l}}\over{\tilde c_s^2}}\label{eqn:mom2}
\end{eqnarray}
 where 
\begin{equation}
{\bf\tilde g_\textrm{eff}} = - {{1}\over{\tilde r^2}} 
{\bf \tilde {\hat r} } 
+ {{1}\over{\tilde R^3}}{\bf \tilde {\hat R} }
\end{equation}
Now in order to obtain a consistent, self-similar solution we require that
the above equation should contain only scaled variables (i.e. those
denoted with a tilde) and  should not depend on the parameters $R_b$, $u_0$ and $c_{so}$ which vary from one simulation to another.
Examining the left hand side of Equation~\ref{eqn:mom2}, we then  require
that $u_0 = c_{s0}$ (i.e. the Mach number of the flow is
the same for the two scaled streamlines at given $\tilde l$).

 We can
furthermore argue that the individual values of $u_0$ and $c_{s0}$
are the same for  streamlines that are a scaled version of each other: if  
$c_{s0}$ were different 
then  the temperature
at given $\tilde l$ would be different between the two streamlines
(by a constant factor) and the density would also be necessarily different according
to the parametrisation of temperature against ionisation parameter. Since  
this relation is not a simple power law scaling then the density variation
along the streamlines would not be self-similar, in contradiction with
our assumption. Therefore, for streamlines to map onto 
one another
in a self-similar way, we must have the {\it same velocity and temperature
structure} as a function of $\tilde l$.

  Turning now to the right hand side of Equation~\ref{eqn:mom2}  we see that in order
that this does not contain variables that differ from case to 
case, we must have $R_{b0} \propto M_*$. Since the escape temperature
is given by 
$\Tesc\propto M_*/r$, this implies that $\Tesc$ is the same at given
$\tilde l$ for streamlines that map onto each other in a self-similar way, thus the importance of $r_g$ as the natural scaling of the problem becomes clear.

\subsection {Primordial discs}

We first consider the case of `primordial' discs on the assumption that
these are approximately scale free. Power law variation of mid-plane 
variables is an expectation of the 
Lynden-Bell \& Pringle (1974) similarity solutions and appear to be
consistent with observational estimates of disc surface density profiles
which suggest that $\Sigma \propto R^{-1}$ (Andrews et al 2010). 
These would therefore appear to be good candidates for  yielding
self-similar solutions when the input parameters are varied, as the
    boundary conditions of the problem would be close to self-similar.

\subsubsection {Variation of $L_X$ at constant $M_*$}

 We have seen above that strict self-similarity requires that
$R_{b0} \propto M_*$. Thus if we vary $L_X$ at  fixed $M_*$, then
$R_{b0}$  
is the same and each streamline simply maps onto
a streamline with the same radius and  with the same temperature and velocity
variation along the streamline. Since the
temperature is a function of ionisation parameter, it follows that 
as one varies $L_X$, the density scaling along each streamline must
simply vary  $ \propto L_X$. The mass flux along each streamline
is the product of stream-bundle area, velocity and density. As the
former two are independent of $L_X$  and
the density scales with $L_X$, then the mass flux is simply
$\propto L_X$. 

 Therefore, we conclude that primordial discs  which differ
only in their values of incident $L_X$ should be topologically identical,  with
identical velocity and temperature structure, but with a mass weighting
on each streamline that scales with $L_X$. This result agrees with
the results of the hydrodynamical simulations of Owen et al (2011b) 
in which it was found that the mass loss
rate scales linearly with $L_X$.

\subsubsection {Variation of $M_*$ at constant $L_X$}

In this case, the fact that $R_{b0}\propto M_*$ implies that variation
of stellar mass should simply change the over-all radial scale of the
flow
(i.e. a streamline
rooted at $10$ A.U. in a simulation with $M= 1 M_\odot$ should simply
map on to a streamline rooted at $1$ A.U. in the case of a simulation
with $M= 0.1 M_\odot$).  As before, we require that the two scaled
streamlines should have the same variation in velocity and temperature
as a function of distance along the streamline normalised to the
radius of the streamline base, i.e. the same variation of
ionisation parameter. Since $\xi= L_X/nr^2$, it follows that
a radial scaling $\propto M_*$ results in  a scaling with density of
$\propto M_*^{-2}$. The mass-loss from the streamline is proportional
to the product of the density ($\propto M_*^{-2}$), the local velocity
(constant) and the area of the stream-bundle ($\propto M_*^2$):
i.e. the mass loss rate along each streamline is the same as the mass
loss rate over its scaled equivalent. Therefore (provided the disc is
sufficiently radially extended that it encompasses the whole region from
which significant mass loss occurs in all cases), we conclude
that, at constant $L_X$, the mass flux in the wind is {\it independent}\footnote {It is of course important to note, that even though there is no explicit mass dependence on photoevaporation rates from discs, there will of course be a strong implicit variation with mass through the variation of X-ray luminosity with stellar mass:  observations suggest that  $M_*\propto L_X^{3/2-5/3}$ (e.g.  Preibisch et al. 2005, Guedel et al. 2007).}
of $M_*$.

\subsection {Discs with inner holes}

 Here the truncated density profile of the disc introduces a particular
radial scale ($R_\textrm{hole}$) to each underlying disc . However, if one fixes $R_{\textrm{hole}}$ for a given mass, then the above arguments for $L_X$  are still applicable. For fixed $M_*$ and $R_{\textrm{hole}}$ then as one varies $L_X$ the density must also vary as $n\propto L_X$,  resulting in a total mass-loss rate that varies linearly with X-ray luminosity. Furthermore, if one varies $M_*$, then one can simply map one solution with an inner hole radius to the equivalent inner hole around another mass using $R_\textrm{hole}\propto M_*$. 

  Neither of these cases however describe the  situation that is normally
of greatest 
interest  i.e. the variation of photoevaporation
rate with $R_{\textrm{hole}}$ in the case of  fixed $M_*$ and
$L_X$, which would allow us to compute  the evolution of the mass loss
rate as a disc is progressively cleared from the inside out. Since $R_{\rm hole}$ introduces a fixed scale into the problem no similarity solution exists than includes $L_X$, $M_*$ and $R_{\rm hole}$ and it is not possible to analytically estimate such a scaling. Therefore, we defer
a discussion of this case to Section ~5 where our numerical results
allow us to construct {\it a posteriori} an approximate scaling argument.

\subsection {Summary}

  We have shown that the flow from primordial discs is expected to
 be self-similar
(when $L_X$ and $M_*$ are varied) and predict that the mass loss rate should
scale linearly with $L_X$ and be independent of $M_*$.  Furthermore, we have shown that photoevaporation rates should scale linearly with $L_X$ also in the
case of discs with inner holes and that a disc with given inner hole
radius can also be scaled to a different hole size ($\propto M_*$) when the 
stellar mass is varied.  The flow topology is however not exactly
self-similar in the case of discs with different hole sizes and fixed $L_X$
and $M_*$. Given the lack of exactly scalable solutions in this
case, we will return to this issue, using additional input from
numerical solutions, in Section 5.

\section{Numerical Fits to Mass-loss Rates}
In this appendix, we provide fits to the total mass-loss rates and
    profiles obtained from the numerical calculations performed by
    Owen et al. (2010), Owen et al. (2011b) and in this work. The
    mass-loss profile fits were based on the results provided by the on-line `function finder' provided at http://zunzun.com, and their functional form does {\it not} represent the results of analytic calculation and should be treated in such a way.  Furthermore, these fits were performed in such a way as to described the cumulative mass-loss rates accurately which is important for the global viscous evolution. As such the fits are worst at small radius; should sensitivity at these radii become important more accurate fits to this region can be provided upon request. The surface mass-loss profiles $\dot{\Sigma}_w$ given in Equations~\ref{eqn:prim_profile} \& \ref{eqn:hole_profile} are provided in normalised form and must be scaled so that $\int2\pi R \dot{\Sigma}_w \dd R$ yields the required total mass-loss rate, given in Equations~\ref{eqn:prim_tot} \& \ref{eqn:hole_tot}. 
\subsection{Primordial Discs}
The total mass-loss rate variation with X-ray luminosity is shown in Figure~\ref{fig:mdot_tot}, where the solid line is described by:
\begin{equation}
\dot{M}_w=6.25\times10^{-9}\left(\frac{M_*}{1\textrm{ M}_\odot}\right)^{-0.068}\left( \frac{L_X}{10^{30} \textrm{erg s}^{-1}}\right)^{1.14}\textrm{\msunyr}\label{eqn:prim_tot}
\end{equation}
The mass-loss profile, derived from the solid line in the left hand panel of Figure~4 in Owen et al. (2011b) is described by:
\begin{eqnarray}
\dot{\Sigma}_w(x>0.7)&\!\!\!\!=\!\!\!\!&10^{(a_1\log_{10}(x)^6+b_1\log_{10}(x)^5+c_1\log_{10}(x)^4)}\nonumber\\&&\times10^{(d_1\log_{10}(x)^3+e_1\log_{10}(x)^2+f_1\log_{10}(x)+g_1)}\nonumber\\&&\times\left(\frac{6a_1\log(x)^5}{x^2\log(10)^7} + \frac{5b_1\log(x)^4}{x^2\log(10)^6} + \frac{4c_1\log(x)^3}{x^2\log(10)^5}\right. \nonumber\\&& + \left.\frac{3d_1\log(x)^2}{x^2\log(10)^4}+\frac{2e_1\log(x)}{x^2\log(10)^3} + \frac{f_1}{x^2\log(10)^2}\right)\nonumber\\ &&\times\exp\left[-\left(\frac{x}{100}\right)^{10}\right]\label{eqn:prim_profile}
\end{eqnarray}
where $a_1=0.15138$, $b_1=-1.2182$, $c_1=3.4046$, $d_1=-3.5717$,
    $e_1=-0.32762$, $f_1=3.6064$, $g_1=-2.4918$ and:
\begin{equation}
x =0.85\left(\frac{R}{\rm AU}\right)\,\left(\frac{M_*}{1\textrm{ M}_\odot}\right)^{-1}
\end{equation}
where $\dot{\Sigma}_w(x<0.7)=0$. Logarithms of the form $\log(x)$ are taken taken using the natural base. 
\subsection{Discs with Inner holes}
As described previously we find that the total mass-loss rate is approximately independent of inner hole radius, which is approximately described by:
\begin{equation}
\dot{M}_w=4.8\times10^{-9}\left(\frac{M_*}{1\textrm{ M}_\odot}\right)^{-0.148}\left( \frac{L_X}{10^{30} \textrm{erg s}^{-1}}\right)^{1.14}\textrm{\msunyr}\label{eqn:hole_tot}
\end{equation}
The mass-loss profile, derived from the solid line in the right hand panel of Figure~4 in Owen et al. (2011b) is described by:
\begin{eqnarray}
\dot{\Sigma}_w(y)&\!\!\!=&\!\!\!\left[\frac{a_2b_2\exp (b_2y)}{R} +\frac{c_2d_2\exp (d_2y)}{R} +\frac{e_2f_2\exp (f_2y)}{R}\right]\nonumber\\
&&\times\exp\left[-\left(\frac{y}{57}\right)^{10}\right]\label{eqn:hole_profile}
\end{eqnarray}
where $a_2=-0.438226$, $b_2=-0.10658387$, $c_2=0.5699464$, $d_2=0.010732277$, $e_2=-0.131809597$, $f_2=-1.32285709$ and:
\begin{equation}
y=0.95\left(R-R_{\rm hole}\right)\left(\frac{M_*}{1\textrm{ M}_\odot}\right)^{-1}
\end{equation}
where $\dot{\Sigma}_w(y<0)=0$. 
\end{appendix}

\label{lastpage}

\end{document}